\documentclass[12pt]{article}

\usepackage{newtxtext,newtxmath}
\usepackage{graphicx}

\usepackage[letterpaper,margin=1in]{geometry}

\renewenvironment{abstract}
	{\quotation}
	{\endquotation}

\date{}

\makeatletter
\renewcommand{\fnum@figure}{\textbf{Figure \thefigure}}
\renewcommand{\fnum@table}{\textbf{Table \thetable}}
\makeatother

\usepackage{scicite}

\usepackage{url}

\def\scititle{
    High-resolution simulations unravel intensification mechanisms of pyrocumulonimbus clouds
}
\title{\bfseries \boldmath \scititle}

\author{
	Qing~Wang$^{1\ast\dagger}$,
	Cenk~Gazen$^{1\dagger}$,
	Matthias~Ihme$^{1,2}$,\and
	Robert~Carver$^{1\ast}$,
	Jeffrey~B.~Parker$^{1}$,
	Tapio~Schneider$^{1,3}$,\and
	Sheide~Chammas$^{1}$,
	Yi-Fan~Chen$^{1}$,
	John~Anderson$^{1}$\and
	\small$^{1}$Google Research, Mountain View, CA 94043, United States.\and
	\small$^{2}$Mechanical Engineering, Stanford University, Stanford, CA 94305, United States.\and
	\small$^{3}$Engineering and Applied Science, California Institute of Technology, Pasadena, CA 91125, United States.\and
	\small$^\ast$Corresponding authors. Email: wqing@google.com, carver@google.com.\and
	\small$^\dagger$These authors contributed equally to this work.
}

\begin{document} 

% Insert the title and author list
\maketitle

% Abstract, in bold
% There are strict length limits, and not all formats have abstracts.
% Consult the journal instructions to authors for details.
% Do not cite any references in the abstract.
\begin{abstract} \bfseries \boldmath
% Start with one or two sentences of background
Pyrocumulonimbus (pyroCb) firestorms---wildfire-generated thunderstorms---can trigger rapid fire spread. However, the multi-physics nature of pyroCb has made their core mechanisms inaccessible to direct observation and previous simulation and prediction efforts.
% Then summarise the results of your observations, experiments, simulations etc.
We introduce a new simulation capability with the first high-resolution, fully coupled simulations of a pyroCb, allowing us to unravel its life cycle governed by two opposing mechanisms. We show fuel moisture is an energy sink that attenuates fire intensity rather than fueling clouds, resolving a long-standing debate. Conversely, we identify the driver of rapid intensification: the Self-Amplifying Fire-Induced Recirculation (SAFIR) mechanism, where precipitation-induced downdrafts intensify the parent fire under weak winds.
% End with a statement of your main conclusions
This work provides a new mechanistic framework for pyroCb prediction and demonstrates a transformative computational approach for previously intractable problems in environmental science.
\end{abstract}

% The first paragraph of any Science paper does NOT have a heading
% Nor is it indented

\noindent
Pyrocumulonimbus (pyroCb) firestorms are among the most extreme manifestations of wildfire, capable of generating their own weather~\cite{FROMM_ETAL_GRL2006}, including lightning and violent surface winds~\cite{Rosenfeld2007-zs,Lareau2016-dh,Tory2018-jr,Fromm2022}, and injecting smoke plumes into the stratosphere, with climatic consequences~\cite{Peterson2018-rj,Ridley2014,Yu2019-cc,KATICH_ETAL_S2023}. These events represent a grand challenge in environmental science, as they emerge from a complex, multi-scale interplay of combustion physics, turbulent fluid dynamics, and cloud microphysics~\cite{TORY_PEACE_THURSTON_BNH2016}. Understanding the mechanisms that govern their behavior is critical for forecasting and mitigating their devastating impacts.

Progress in understanding these events has been fundamentally limited by both observational and computational barriers~\cite{TORY_PEACE_THURSTON_BNH2016}. Direct, comprehensive observation of the fire-atmosphere feedback loops within a pyroCb is prohibitively dangerous and difficult~\cite{Peterson2022-wm}. Consequently, scientific understanding has relied on numerical models~\cite{Trentmann2006-qz,Luderer2006-ni,Reutter2014-mm,Thurston2016-om,Badlan2021-ds,Peace2023-ic,Lee2023-nd,Turney2023-gm}. However, resolving key debates, such as the role of fuel moisture in cloud formation~\cite{TORY_PEACE_THURSTON_BNH2016,Cunningham2009-nv,Luderer2009-qb}, has been impossible because existing simulation approaches have lacked the necessary resolution or the fully coupled fire-atmosphere physics to capture these critical interactions. Without the ability to simulate the event in its entirety, the complete life cycle of a pyroCb has remained uncharacterized~\cite{Trentmann2006-qz,Luderer2006-ni,Reutter2014-mm,Thurston2016-om,Badlan2021-ds,Peace2023-ic,Lee2023-nd, Reisner2023-gp, Turney2023-gm}.

To overcome these barriers, we conducted the first high-resolution, fully coupled fire-atmosphere simulations of a pyroCb, leveraging a new generation of high-fidelity computational tools~\cite{Wang2022-ln,Wang2023-uv,Chammas2023-ts} capable of modeling these events at an unprecedented scale and physical fidelity. This predictive simulation capability enables scientists, for the first time, to analyze the complete system and quantify the competing mechanisms that govern the pyroCb life cycle from initiation to intensification or decay. Our simulations reveal that pyroCb evolution is dictated by a balance between two distinct, opposing pathways: a moisture-driven attenuation that acts as a brake on the system, and a downdraft-driven self-intensification (the SAFIR mechanism) that acts as an accelerator.

This study quantitatively resolves the disputed role of fuel moisture and provides the first detailed representation of the feedback loop that drives eruptive fire growth in low-wind environments. More broadly, it demonstrates the power of high-fidelity, physics-based simulation to deconstruct complex environmental systems, offering critical insights for forecasting the growing threat posed by extreme fire-weather events.

% Research Articles and Reviews split the text into sections using headings
% Use a short (up 6 words) descriptive phrase, not generic 'Results' or 'Conclusions'
% Most other formats do not have headings, see the journal instructions to authors for details
\subsection*{\label{SEC_MODEL}Model configuration and case study}

To investigate the mechanisms governing pyrocumulonimbus (pyroCb) development, we conducted a series of high-resolution, fully coupled fire-atmosphere simulations using a large-eddy simulation (LES) framework~\cite{Wang2022-ln}. Detailed model equations, physical conditions, and simulation configurations are provided in materials and methods.

The baseline simulation approximates the fuel and atmospheric conditions of the 2019 Williams Flat fire~\cite{Peterson2022-wm}, which produced notable pyroCb events (Fig.~\ref{fig:baseline}A). We chose to simulate the fire-cloud system using a horizontal grid spacing of 5 m in the area of the fire and a vertical spacing of 0.5 m, stretching to 190 m at the model top (20 km). This fine resolution over such a tall domain combined with a physics-based combustion model representing fire processes is novel for coupled fire-atmosphere simulations.

To explore the sensitivity of pyroCb development and to illustrate the underlying physical mechanisms, we conducted three additional simulations where key parameters were perturbed relative to the baseline:
\begin{itemize}
    \item Reduced Wind Speed (1/3 Wind): The mean ambient wind profile was scaled by a factor of 1/3. This case explores the system's response under significantly weaker wind shear.
    \item Reduced Wind Speed (2/3 Wind): The mean ambient wind profile was scaled by a factor of 2/3. This case extends the 1/3 wind case to quantify the sensitivity of the wind shear on the fire-pyroCb interaction.
    \item Increased Fuel Moisture (30\% MC): Fuel moisture content (MC) was increased from 0.01\% to 30\%. This perturbation was chosen to investigate the significant, yet often debated, impact of fuel moisture on fire behavior and pyroCb formation, over a range relevant for transitioning from dry to moderately moist fuel beds.
\end{itemize}

\begin{figure}
    % \centering is crucial to center the entire figure content on the page.
    \centering
    \includegraphics[width=0.9\textwidth]{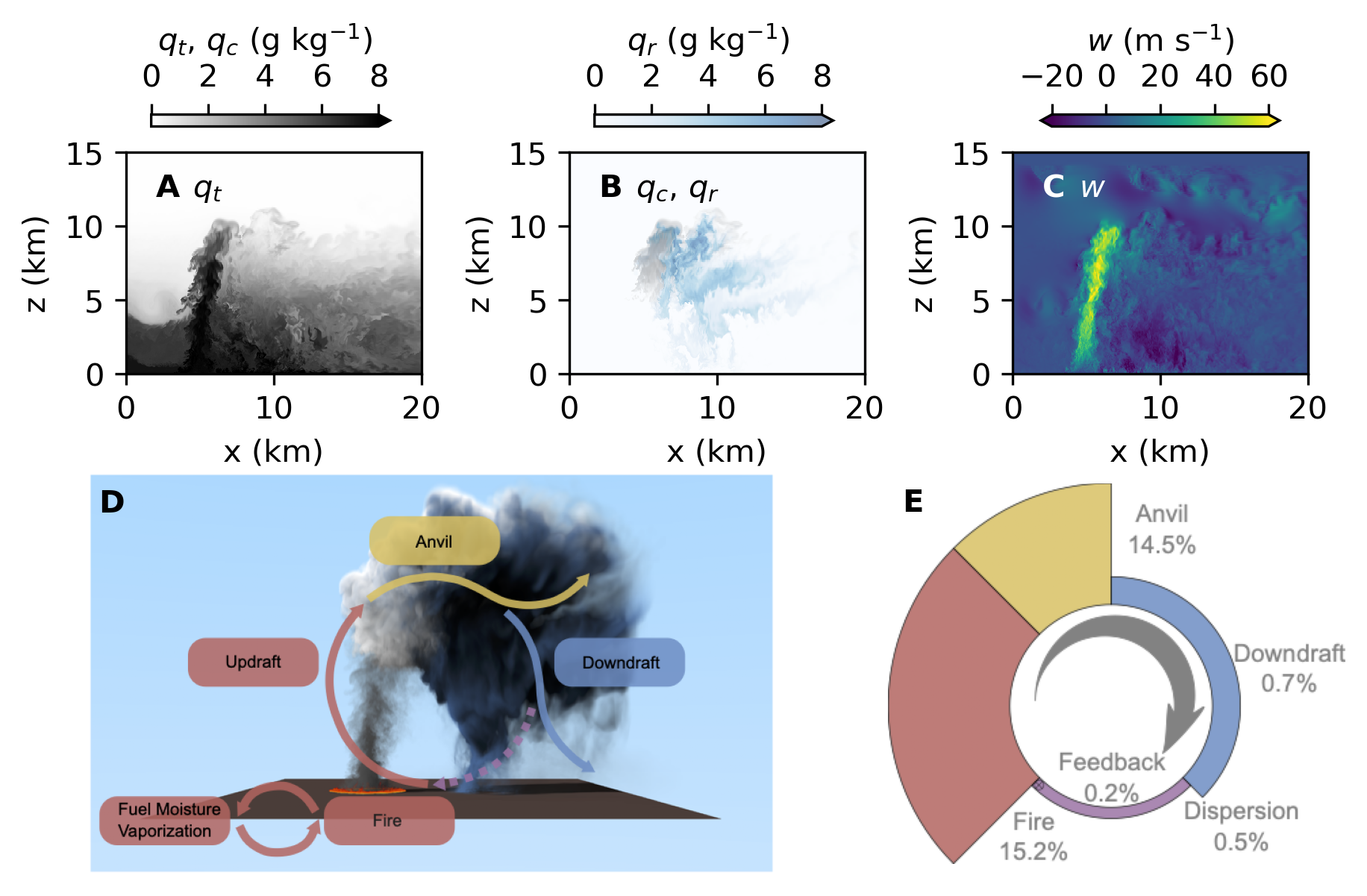}
    \caption{\textbf{Conceptual model and quantitative balances of baseline simulations.} \textbf{A}, Center plane cross-section of total humidity. \textbf{B}, Center plane cross-section of cloud (black) and rain (blue) fractions. \textbf{C}, Center plane cross-section of the vertical velocity. \textbf{D}, A conceptual model illustrating the five core processes influencing a pyroCb event: fire-driven convection, vaporization, updraft, downdraft, and dispersal into the anvil. The purple dashed branch represents a potential feedback from the downdraft to the fire, which is more pronounced under a low wind condition. \textbf{E}, The fraction of air parcels that pass through each process in the conceptual model for the baseline simulation at 1 hour and 22 minutes.}
    \label{fig:baseline_model}
\end{figure}

Figure~\ref{fig:baseline_model} illustrates key aspects of this simulation at 2 hours post-ignition. Under the low ambient wind-shear conditions characteristic of this event, the fire plume rises vertically (Figs.~\ref{fig:baseline_model}A and~\ref{fig:baseline_model}B), driven by strong buoyancy from the fire's heat release, with peak updraft velocities reaching 60 m/s (Fig.~\ref{fig:baseline_model}C). An anvil-shaped cloud forms above the burn area, with its top extending to the upper troposphere at approximately 11 km. Significant rainfall is observed, primarily as virga, with a rain-induced downdraft carrying some combustion products towards the ground, while a substantial portion remains aloft, dispersing downstream (Fig.~\ref{fig:baseline}A, movie S1). 

Distinct behaviors, including fire whirls and rapid fire spread driven by wind, underscore the strong coupling between the fire and the atmosphere (Figs.~\ref{fig:baseline}B and~\ref{fig:3d_rendering}, movies S2 and S3). We conceptualize the pyroCb event with five core processes (Fig.~\ref{fig:baseline_model}D). The fire acts as the primary heat source inducing convection. The vaporization of fuel moisture serves dually as a potential water source for the atmosphere and a significant energy sink for the fire. Convection transports heat and mass aloft, forming an updraft plume. Detrainment of precipitation from the updraft induces downdrafts. Critically, air from these downdrafts can recirculate back into the fire, establishing a direct interaction pathway.

To quantify the dynamics within these processes, we performed a detailed
Lagrangian parcel tracking study, using $10^5$ passive tracer parcels.  The subsequent analysis of parcel fates (Fig.~\ref{fig:baseline_model}E and~\ref{fig:parcel_count_sankey}) and thermodynamic evolution (Supplementary Text~\ref{SUP_SEC_PARCEL}) forms the basis for identifying and quantifying the mechanisms discussed in what follows.

% If your text is very short you might need to uncomment the following line to avoid
% layout problems with the figures and tables.
%\newpage

\subsection*{\label{SEC_ATTENUATION}The Attenuation Mechanism: Fuel Moisture as an Energy Sink}

Our simulations reveal fuel moisture content (MC) attenuates fire and pyroCb, predominantly by acting as a significant surface energy sink rather than a substantial direct moisture source for the pyroCb cloud. This attenuation pathway directly impacts the available combustion heat for convection, a critical factor highlighted in our conceptual model (Fig.~\ref{fig:baseline_model}D).

The parcel budget analyses highlight this effect (Figs.~\ref{fig:energy_budget}C and~\ref{fig:energy_budget_ext}). The energy budget for air parcels near the surface (fig.~\ref{fig:energy_budget_ext}) demonstrates that while combustion is the major heat source, vaporization for fuel moisture represents a dominant heat sink. In our 30\% MC simulation, the fire is suppressed through two primary effects: first, 2.2\% of the combustion energy is directly consumed by vaporization (Figs.~\ref{fig:energy_budget}C and~\ref{fig:energy_budget_ext}D); second, the associated cooling of the fuel bed further reduces the rate of combustion. The combined result is a 37.4\% reduction in total fire power (Figs.~\ref{fig:energy_budget}C and~\ref{fig:energy_budget_ext}E). This direct energy loss results in a quantifiable reduction in the available fire power and suppresses the rate of fire spread compared to the dry fuel baseline. Consequently, the energy available to drive convective updrafts is diminished, as evidenced by the reduced potential energy budget for parcels in the 30\% MC case (Figs.~\ref{fig:energy_budget}C and~\ref{fig:energy_budget_ext}A).

This reduction in available energy directly translates to a less buoyant and less vigorous fire plume. Thermodynamic analysis of mean updraft parcel properties (table~\ref{tab:skew_t_parcel}, fig.~\ref{fig:skew_t_parcel}) clearly illustrates this: the 30\% MC simulation exhibits a markedly smaller fireCAPE~\cite{Tory2018-jr} of 1921 J/kg compared to 2899 J/kg in the dry fuel baseline. This reduced fireCAPE signifies less available buoyant energy for parcel ascent. Consequently, the theoretical maximum vertical velocity ($w_{\max}$) is significantly lower in the 30\% MC case (62.0 m/s) compared to the baseline (76.2 m/s)—a reduction of approximately 18.6\%. This reduction in updraft strength indicates a direct suppression of the plume's ability to reach higher altitudes and sustain vigorous convection. Furthermore, the drier fuel case resulted in a less diluted plume at its top, as its increased buoyancy allowed it to ascend more rapidly, reducing entrainment and suggesting drier fuels lead to more vigorous plumes capable of reaching the stratosphere.

While fuel moisture critically dampens fire intensity and plume buoyancy by consuming energy, our findings confirm it is a minor contributor of water mass to the actual pyroCb cloud. To quantify water sources, we traced parcels backward from the cloud region (parcels with $q_c>1$ g/kg) to their origins (Fig.~\ref{fig:baseline}C). The results, summarized in Table.~\ref{tab:water_contribution}, show that for all simulation cases, lateral entrainment of ambient environmental air is the dominant source of water for the pyroCb, accounting for approximately 90-97\% of the total cloud water---similar to regular cumulonimbus clouds \cite{Schiro18a}. Combustion products contribute a smaller fraction (roughly 2-10\%, Figs.~\ref{fig:energy_budget}B and~\ref{fig:water_budget_ext}D), while fuel moisture vaporization contributes less than 1\% (Figs.~\ref{fig:energy_budget}B and~\ref{fig:water_budget_ext}C), even in the 30\% MC case. These findings are consistent with, and lend further support to, prior studies~\cite{Luderer2009-qb} suggesting fire-released moisture has a minimal direct impact on pyroCb cloud water content, especially for fires in deep, well-mixed boundary layers where substantial plume dilution occurs.

\subsection*{\label{SEC_INTENSIFICATION}The Self-Intensification Mechanism: The SAFIR Feedback Loop}

In contrast to the attenuation driven by fuel moisture, our simulations reveal a potent self-intensification pathway that can lead to rapid and extreme fire growth. This second key factor, which we term the self-amplifying fire-induced recirculation (SAFIR) mechanism, is driven by a positive feedback between the precipitation-induced downdraft and the fire itself.

\begin{figure}
  \centering
  \includegraphics[width=0.9\textwidth]{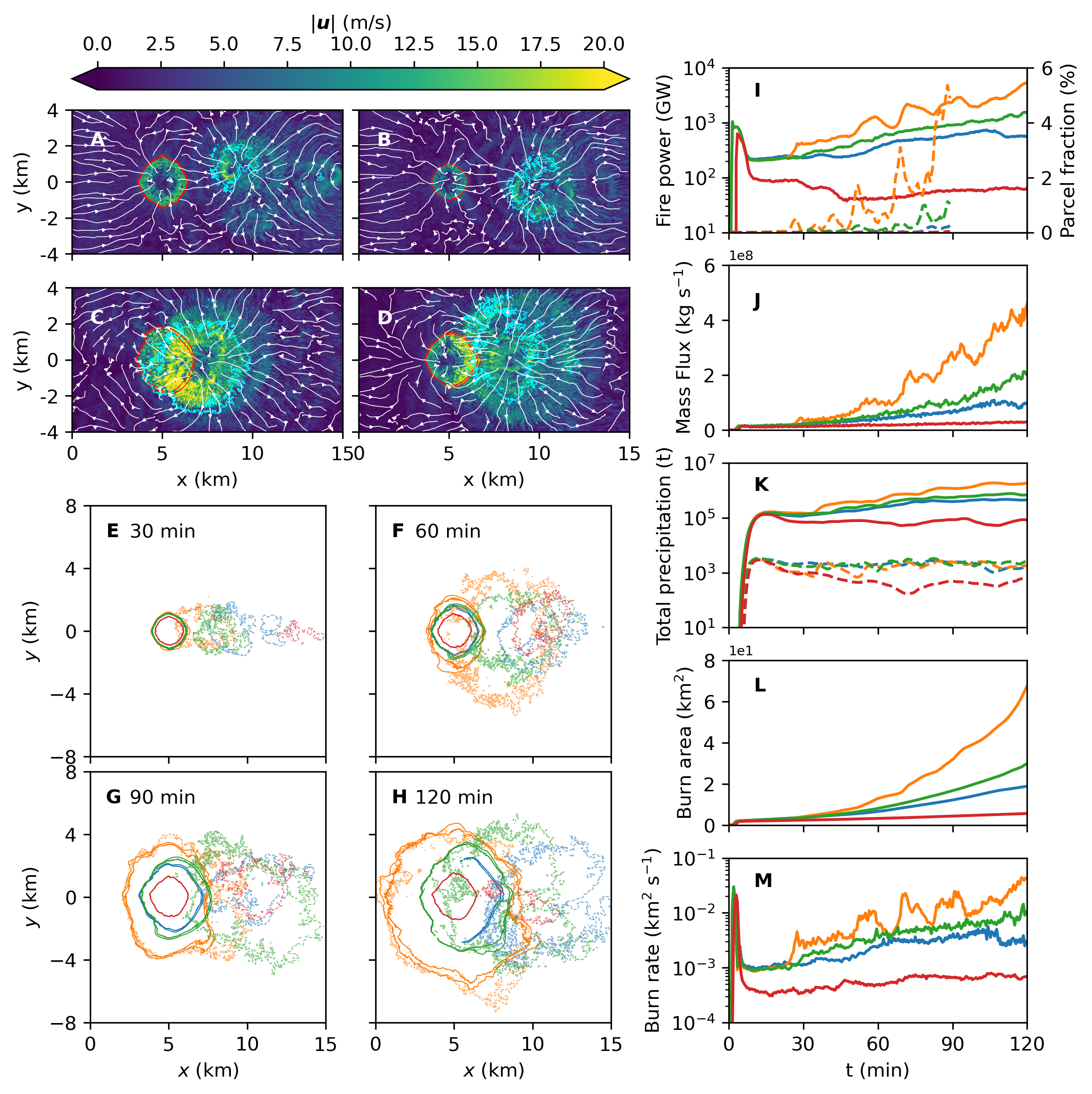}
  \caption{\textbf{Simulation statistics illustrating the SAFIR mechanism.} Comparison of the four simulation configurations (blue: baseline; orange: 1/3 wind; green: 2/3 wind; red: 30\% MC). \textbf{A-D}, Velocity magnitude and streamlines at 12 m altitude at 52 minutes for the baseline (\textbf{A}), 30\% MC (\textbf{B}), 1/3 wind (\textbf{C}), and 2/3 wind (\textbf{D}) cases, respectively. Red contours show the fire perimeter ($T=600$ K); cyan contours show the rain front. \textbf{E-H}, Time evolution of the fire perimeter (solid lines) and rain front (dotted lines) for all cases at 30 (\textbf{E}), 60 (\textbf{F}), 90 (\textbf{G}), and 120 (\textbf{H}) minutes, respectively. \textbf{I}, Fire power (solid lines) and fraction of recirculated air parcels (dashed lines). \textbf{J}, Mass flux at 1 km above the ground level. \textbf{K}, Total accumulated precipitation. \textbf{L}, Total burned area. \textbf{M}, Burning rate.}
  \label{fig:fire_atmos_stats}
\end{figure}

The direct consequence of this feedback is a dramatic amplification of fire power. In the 1/3 wind case, the onset of significant parcel recirculation (fig.~\ref{fig:parcel_count_sankey}) directly precedes a surge in fire power, which increases to a mean of 1907 GW, roughly three times the 481 GW mean of the baseline case (Fig.~\ref{fig:fire_atmos_stats}I). This tripling of fire power is associated with a 30-fold increase in the fraction of recirculated parcels (from 0.039\% to 1.18\%), confirming the profound impact of the fire-downdraft interaction.

\begin{figure}
    \centering
    \includegraphics[width=0.9\textwidth]{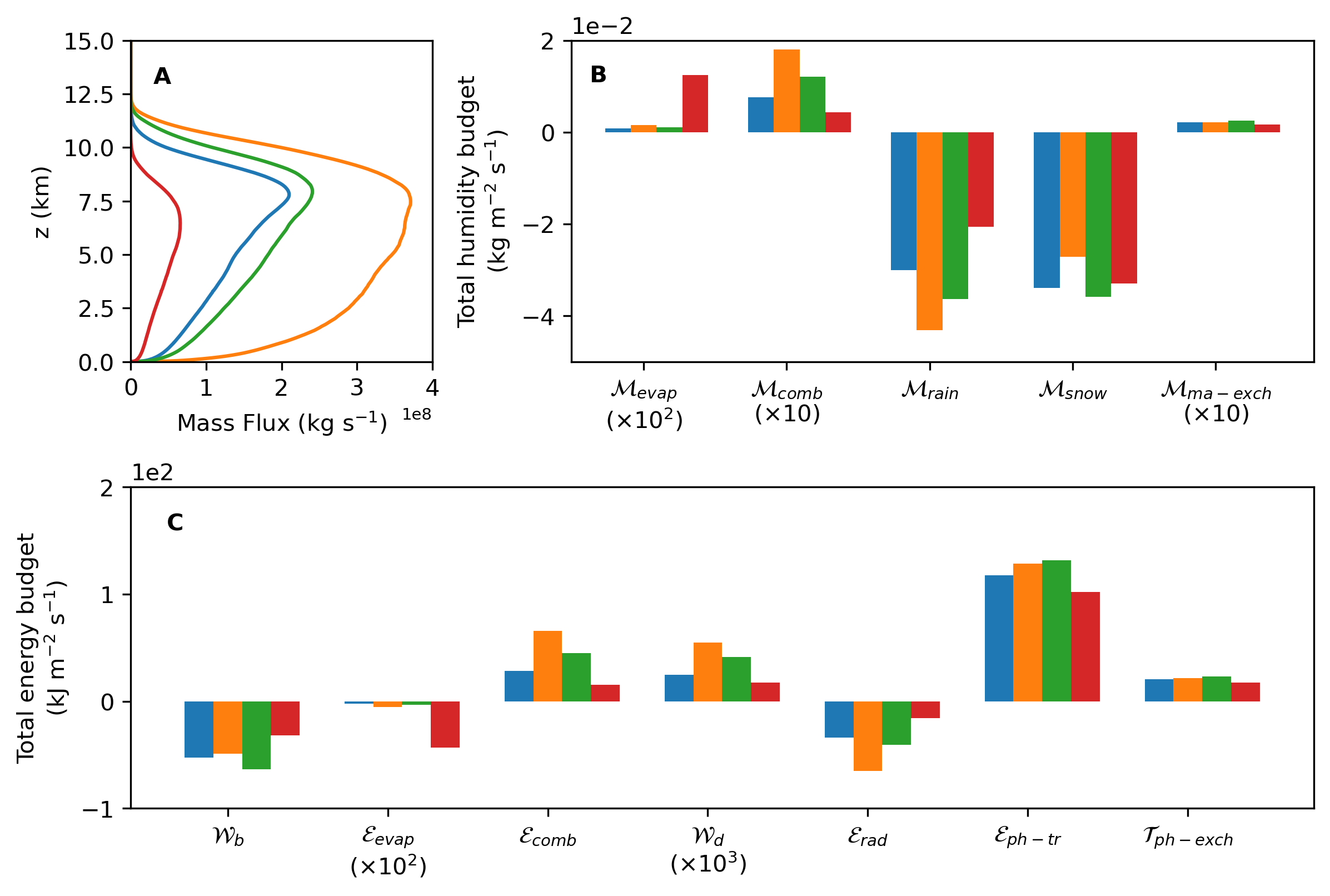}
    \caption{\textbf{Parcel analyses of key physical processes in the plume.} Comparison of the four simulation configurations (blue: baseline; orange: 1/3 wind; green: 2/3 wind; red: 30\% MC). \textbf{A}, The mean mass flux of air from 20 to 120 minutes. \textbf{B}, The vertical integration of the mean humidity budget terms that are computed based on eq.~\ref{eq:q_t_parcel}. Groups from left to right represent contributions from vaporization of fuel moisture ($\mathcal{M}_\text{evap}$), combustion product ($\mathcal{M}_\text{comb}$), rain ($\mathcal{M}_\text{rain}$), snow ($\mathcal{M}_\text{snow}$), and moisture transfer due to mass exchange ($\mathcal{M}_\text{ma-exch}$), respectively. \textbf{C}, The vertical integration of the mean energy budget terms that are computed based on eq.~\ref{eq:total_energy_density_parcel_with_solid}. Groups from left to right represent contributions from buoyancy work ($\mathcal{W}_b$), fuel-moisture vaporization ($\mathcal{E}_\text{evap}$), combustion heat release ($\mathcal{E}_\text{comb}$), canopy-drag heating ($\mathcal{W}_d$), radiative heat flux ($\mathcal{E}_\text{rad}$), latent heat release from precipitation condensation ($\mathcal{T}_\text{ph-tr}$), heat transfer due to mass exchange ($\mathcal{T}_\text{ph-exch}$), respectively.}
    \label{fig:energy_budget}
\end{figure}

The SAFIR mechanism unfolds in a distinct sequence:
\begin{itemize}
    \item Downdraft-Surface Interaction: Precipitation from the pyroCb induces a strong downdraft. As this downdraft impinges on the ground, it generates large-scale horizontal vortices, accelerating a lateral near-surface flow radially outward from the point of impact that reaches 10 to 20 m/s (as shown by streamlines in Figs.~\ref{fig:fire_atmos_stats}A to~\ref{fig:fire_atmos_stats}D and the full time sequence in fig.~\ref{fig:ground}).
    \item Enhanced Fire Inflow: The lateral near-surface flow interacts with the fire environment, creating a counter-flow to the ambient wind on the upwind side and reinforcing it on the downwind side. This interaction critically enhances surface inflow into the fire perimeter from multiple directions, accelerating inflow streams from less than 5 m/s (Figs.~\ref{fig:fire_atmos_stats}A and~\ref{fig:fire_atmos_stats}B) to over 20 m/s when the downdraft is in close vicinity of the fire (Figs.~\ref{fig:fire_atmos_stats}C and~\ref{fig:fire_atmos_stats}D).
    \item Fire Intensification: The enhanced inflow dramatically increases the fire's rate of spread (fig.~\ref{fig:fire_front_and_ros}) and total energy release, as quantified by the surge in fire power (Fig.~\ref{fig:fire_atmos_stats}I), total burned area (Fig.~\ref{fig:fire_atmos_stats}L),  burning rate (Fig.~\ref{fig:fire_atmos_stats}M).
    \item Feedback Amplification: The intensified fire drives a more vigorous convective updraft (Figs.~\ref{fig:energy_budget}A and~\ref{fig:fire_atmos_stats}J). This stronger updraft leads to a more intense pyroCb, which in turn can produce more substantial precipitation and stronger downdrafts (Fig.~\ref{fig:fire_atmos_stats}K), closing the feedback loop.
\end{itemize}

The activation of the SAFIR mechanism is sensitive to environmental conditions, particularly ambient wind speed. In the low-wind-speed simulation (1/3 wind), the precipitation and its associated downdraft are not advected far from the fire, allowing for an earlier and more sustained interaction with the fire perimeter (Figs.~\ref{fig:fire_atmos_stats}E to~\ref{fig:fire_atmos_stats}H). This is important because the peak surface downdraft winds are found shortly after the downdraft reaches the surface~\cite{Orf2012-bl}.  Specifically, while the feedback cycle is first observed ${\sim}79$ minutes after ignition in the baseline configuration, this time reduces to 50 and 24 minutes for the 2/3 and 1/3 wind cases, respectively. This proximity is critical for initiating the feedback loop. In contrast, in cases with higher wind speeds, the downdraft is advected further downstream, preventing significant interaction. Similarly, in the moisture-attenuated case (30\% MC), the fire lacks the intensity to produce a pyroCb strong enough to generate the significant precipitation and downdrafts needed to trigger the mechanism, resulting in a steady, predictable rate of fire spread.

This downdraft-fire feedback also has a profound impact on the local weather conditions generated by the pyroCb, driving more intense and pulsating convection (movie S3). This is reflected in the total precipitation, which increases substantially under the low-wind condition in proportion to the enhanced fire power (Fig.~\ref{fig:fire_atmos_stats}K). This non-linear dynamic, where a downdraft ultimately leads to a more intense storm, highlights a behavior counterintuitive (because it preferentially occurs under low ambient wind) but critical for understanding extreme fire events. The abrupt and rapid fire spread resulting from the SAFIR feedback loop represents an exceptionally hazardous condition that is not captured by classical fire spread models.

\subsection*{\label{SEC_DISC}Discussion}

By employing high-fidelity, fully coupled fire-atmosphere simulations, this study has quantitatively analyzed the competing mechanisms governing the life cycle of pyroCb events. Our results move beyond generalized descriptions of fire-atmosphere interactions to identify and characterize two distinct, opposing pathways: a moisture-driven attenuation and a downdraft-driven self-intensification, with the prevailing environmental conditions dictating which mechanism dominates.

Our analysis resolves the long-standing debate on the role of fuel moisture in pyroCb dynamics, demonstrating quantitatively its primary role as an energy sink that significantly attenuates the fire. By consuming energy that would otherwise contribute to combustion, high fuel moisture suppresses fire power and diminishes the available potential energy (fireCAPE) for convection. This initiates an attenuation cascade: the less energetic fire drives a weaker, less buoyant updraft, which reduces the entrainment of ambient air and ultimately results in a smaller, less severe pyroCb. Furthermore, we confirm moisture from fuel is a negligible source ($<$1\%) for the pyroCb cloud's water mass, which is instead dominated by entrainment of ambient air.

The most critical discovery is the identification and characterization of the self-amplifying fire-induced recirculation (SAFIR) mechanism, a feedback loop where precipitation-induced downdrafts intensify the parent fire. Our simulations provide the first detailed representation of this process. The strong positive correlation between recirculated parcels and the non-linear surge in fire power provides convincing evidence for its efficacy. The novelty of SAFIR lies in its demonstration of a pathway to extreme fire behavior found in low ambient wind speeds---a sharp contrast to classical models where fire intensity scales with wind speed \cite{Andrews2018-ms}. This reveals a mechanism for sudden, eruptive fire growth that could be missed by conventional forecasting methods, which often associate the highest danger with strong winds.

The activation of these competing mechanisms is dictated by the environmental context. The SAFIR feedback loop was triggered only under low-wind conditions, when the downdraft remained in close proximity to the fire. Higher wind speeds advected the downdraft downstream, preventing the crucial interaction. This highlights a key insight: while strong winds are traditionally associated with dangerous fire spread, low-wind environments may harbor the potential for a different mode of extreme, unpredictable intensification if a pyroCb can form and initiate the SAFIR loop.

Finally, our results underscore the power of high-fidelity, physics-based simulations. Tracing energy and mass pathways with Lagrangian particle tracking through the coupled system was essential for identifying these mechanisms. The balance between attenuation and SAFIR-driven intensification controls pyroCb intensity, which in turn determines the quantity of smoke and pollutants injected aloft. Given the significant climatic and environmental implications of these injections, understanding these fundamental mechanisms is a critical step toward improving both near-field hazard prediction and long-range atmospheric impact assessments.

%TC:ignore

%%%%%%%%%%%%%%%% MAIN TEXT FIGURES %%%%%%%%%%%%%%%

\begin{figure} % Do NOT use \begin{figure*}
    % \centering is crucial to center the entire figure content on the page.
    \centering
    \includegraphics[width=0.9\textwidth]{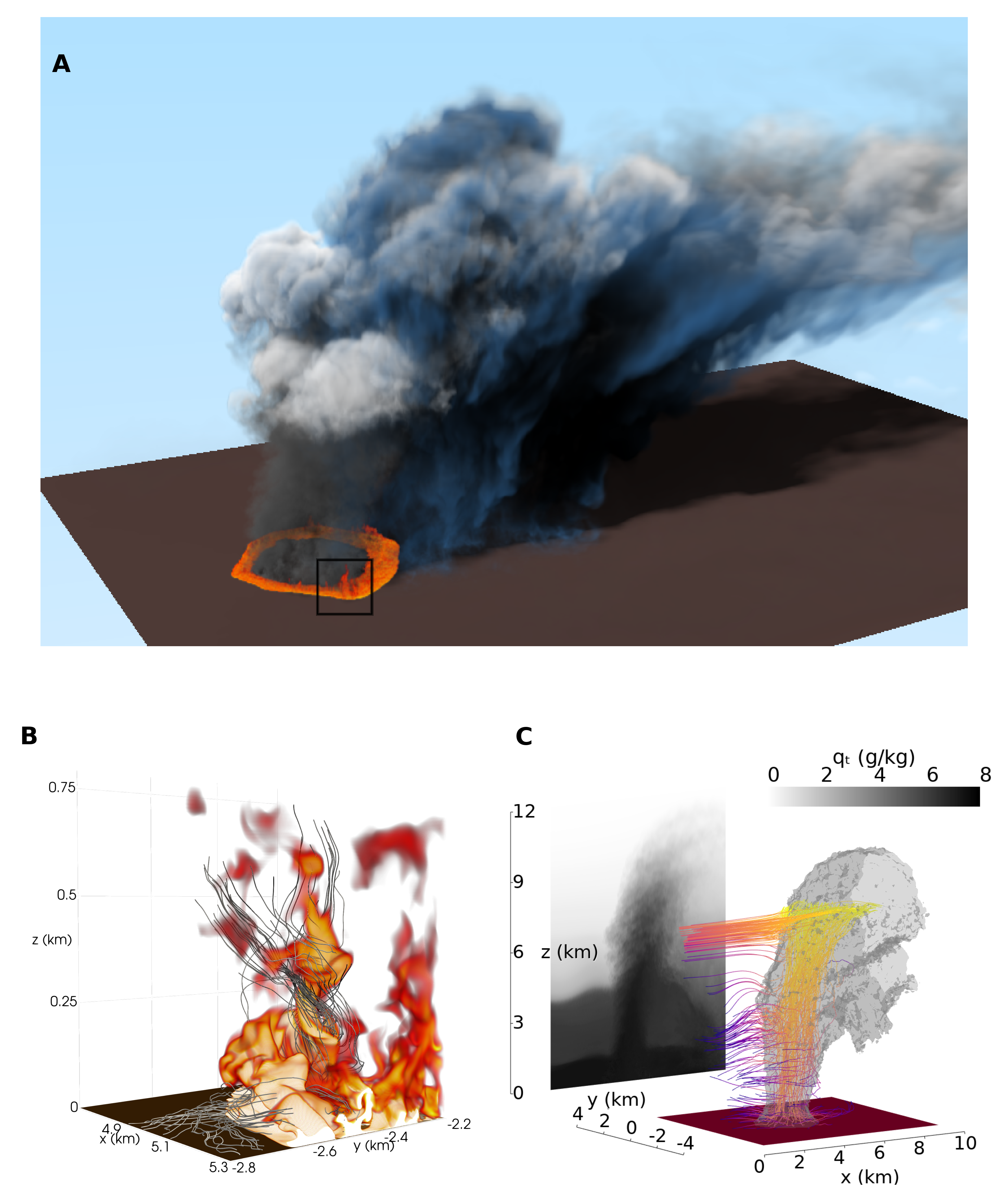} % for an image file named example_figure.*
    % Pick an appropriate width - in print, figures are usually one or two columns wide, which can
    % be approximated by 0.3\textwidth or 0.6\textwidth respectively. Use appropriate label sizes.
    
    % Captions go below figures
    \caption{\textbf{Illustration of baseline simulation dynamics.} \textbf{A}, 3D rendering of the fire (orange), smoke (grey), cloud (white), and rain (blue) at 2 hours post-ignition. \textbf{B}, A fire whirl with tracer air parcels (thin black lines) within the highlighted region in a. \textbf{C}, The mean plume with tracer air parcels (colored lines).}
    \label{fig:baseline} % give each figure a logical label name
\end{figure}

%%%%%%%%%%%%%%%% MAIN TEXT TABLES %%%%%%%%%%%%%%%

\begin{table} % Do NOT use \begin{table*}
    \centering
    \caption{\textbf{Fractional contribution of water sources to pyroCb cloud formation.} The table shows the percentage contribution of water from three different sources (ambient environmental entrainment, water released from combustion, and fuel moisture vaporization) to the total mass of the pyroCb cloud for each of the four simulation cases. For all cases, lateral entrainment of ambient air is the dominant source ($>$90\%), while vaporization from fuel moisture contributes less than 1\%, even in the 30\% MC case.}
    \label{tab:water_contribution}
    \begin{tabular}{ccccc}
        \\
        \hline
        & Baseline & 1/3 Wind Speed & 2/3 Wind Speed & 30\% MC \\
        \hline
        Entrainment & 94.8 & 90.2 & 93.4 & 97.2 \\
        Combustion & 5.1 & 9.6 & 6.6 & 2.1 \\
        Vaporization & 0.1 & 0.1 & 0.1 & 0.7 \\
        \hline    
    \end{tabular}
\end{table}

%%%%%%%%%%%%%%%% REFERENCES %%%%%%%%%%%%%%%

\clearpage % Clear all remaining figures and tables then start a new page

% The list of references goes after the main text and before the acknowledgements
% When preparing an initial submission, we recommend you use BibTeX, like this:
%
\bibliography{references} % for a file named references.bib
\bibliographystyle{sciencemag}

% After the paper has completed peer review and been revised ready for acceptance,
% you should comment out the lines above and copy-paste the contents of your .bbl
% file here instead. This will help ensure that our conversion software works correctly.
% Remember to re-run BibTeX first - check the timestamp!
%
% Example of the first three entries copy-pasted from science_template.bbl:
%
%\begin{thebibliography}{1}
%
%\bibitem{example}
%A.~N. {Author}, An example reference. \emph{Journal of Improbable Research}
%  \textbf{1}, 67 (2020).
%
%\bibitem{example2}
%F.~M. {Surname}, S.~{Author}, A second example. \emph{Interesting Research
%  Letters} \textbf{32}, 897 (2019).
%
%\bibitem{example_preprint}
%P.~{One}, P.~{Two}, P.~{Three}, {An unpublished preprint}. \emph{preprint}
%  (2021), arXiv:2101.12345.
%
%\end{thebibliography}

%%%%%%%%%%%%%%%% ACKNOWLEDGEMENTS %%%%%%%%%%%%%%%

\section*{Acknowledgments}
We thank Anudhyan Boral, James Lotte, Rasmus Larsen, and Sameer Agarwal for technical help in the LES development; Fei Sha for his editorial suggestions; Tyler Russell and Carla Bromberg for technical program management; and Corinna Cortes, Yossi Matias, Fernando Pereira, and John Platt for leadership support.

\paragraph*{Funding:}
This work is funded by Alphabet Inc.

\paragraph*{Author contributions:}
\textbf{Conceptualization}: Q.W., C.G., M.I., R.C., T.S., Y.-F.C., J.A.
\textbf{Methodology}: Q.W., M.I.
\textbf{Software}: Q.W., C.G., M.I., J.B.P., S.C., T.S., Y.-F.C.
\textbf{Validation}: Q.W., C.G., M.I., R.C., 
\textbf{Formal analysis}: Q.W., C.G., R.C., M.I.
\textbf{Investigation}: Q.W., C.G., M.I., R.C., T.S., Y.-F.C., J.A.
\textbf{Resources}: Y.-F.C., J.A.
\textbf{Data Curation}: Q.W., C.G.
\textbf{Writing---Original Draft}: Q.W., C.G., R.C., M.I.
\textbf{Writing---Review and Editing}: Q.W., R.C., T.S., C.G., M.I., Y.-F.C.
\textbf{Visualization}: Q.W., C.G., R.C.
\textbf{Project Administration}: Y.-F.C.
\textbf{Supervision}: Y.-F.C., J.A.

\paragraph*{Competing interests:}
All authors are employees of Google LLC and own Alphabet stock as part of the standard compensation package.

\paragraph*{Data and materials availability:}
Simulation results are available at
\url{https://console.cloud.google.com/storage/browser/pyrocb}. The code needed to reproduce the results is available at \url{https://github.com/google-research/swirl-lm}.

%%%%%%%%%%%%%%%% SUPPLEMENT LIST %%%%%%%%%%%%%%%

% List the contents of your Supplementary Materials, including the numbers of any
% supplementary figures, tables, external data files etc. and any references that are
% cited only in the supplement. In this example, refs. 7-8 are cited only in the supplement.
% Fill out your numbers accordingly and delete any lines that aren't applicable.
\subsection*{Supplementary materials}
Materials and Methods\\
Supplementary Text\\
Figs. S1 to S9\\
Table S1\\
References \textit{(29-\arabic{enumiv})}\\ % automatically fills out the last reference number
% (filling out the other numbers automatically is possible but fiddly and liable to break)
Movies S1 to S3\\

%%%%%%%%%%%%%%%% END OF MAIN TEXT %%%%%%%%%%%%%%%

\newpage

%%%%%%%%%%%%%%%% START OF SUPPLEMENT %%%%%%%%%%%%%%%

% Figures, tables, equations and pages in the supplement are numbered S1, S2 etc.
\renewcommand{\thefigure}{S\arabic{figure}}
\renewcommand{\thetable}{S\arabic{table}}
\renewcommand{\theequation}{S\arabic{equation}}
\renewcommand{\thepage}{S\arabic{page}}
\setcounter{figure}{0}
\setcounter{table}{0}
\setcounter{equation}{0}
\setcounter{page}{1} % not 0 as \newpage already started a supplementary page
% References continue the numbering from the main text.

%%%%%%%%%%%%%%%% SUPPLEMENT TITLE PAGE %%%%%%%%%%%%%%%

\begin{center}
\section*{Supplementary Materials for\\ \scititle}

% Author list for the supplement
% Indicate the corresponding authors, but do NOT include institutions here
% It would be nice if the template auto-generated this, but doing so is complicated...
Qing~Wang$^{\ast\dagger}$,
Cenk~Gazen$^{\dagger}$,
Matthias~Ihme,\\
Robert~Carver$^{\ast}$,
Jeffery~B.~Parker,
Tapio~Schneider,\\
Sheide~Chammas,
Yi-Fan~Chen,
John~Anderson\\ % we're not in a \author{} environment this time, so use \\ for a new line
\small$^\ast$Corresponding authors. Email: wqing@mail.com, carver@google.com\\
\small$^\dagger$These authors contributed equally to this work.
\end{center}

% Fill out the numbers for each type of supplementary material,
% and delete any lines that aren't applicable.
% These are just example numbers that don't match the rest of this template.
\subsubsection*{This PDF file includes:}
Materials and Methods\\
Supplementary Text\\
Figures S1 to S9\\
Table S1\\
Captions for Movies S1 to S3

\subsubsection*{Other Supplementary Materials for this manuscript:}
Movies S1 to S3

\newpage

%%%%%%%%%%%%%%%% MATERIALS AND METHODS %%%%%%%%%%%%%%%

\subsection*{Materials and Methods}

\subsubsection*{\label{SUB_SEC_CONFIG}Simulation Configurations}

All pyroCb simulations in this study are performed on a domain of size $30\times 20\times 20\ \text{km}^3$. The mesh is stretched in all 3 dimensions. In the $x$ (along-wind) dimension, the mesh is kept uniform at 5 m resolution from 0 to 7.5 km, and stretched linearly with a ratio of 1.007 to 30 km. In the $y$ (across-wind) dimension, the mesh resolution is kept at 5 m between $-4$ km and $4$ km, and stretched with a ratio of 1.014 to 10 km on both ends. In the $z$ (vertical) dimension, the mesh resolution is 0.5 m at the bottom and stretched linearly with a ratio of 1.006 to 2 km, followed by a ratio of 1.004 to 20 km, which results in a largest grid spacing of 190 m at 20 km. The domain is partitioned across 256 TPU v5e cores, with a topology of $8\times 8\times 4$ along the $x$, $y$, and $z$ dimensions. Each core has a sub-mesh of size $256\times 256\times 128$ points, resulting in a total of 2.15 billion grid points. Each simulation is performed for 2 physical hours post-ignition, with an average wall-clock time of 2 days.

\begin{figure} % Do not use \begin{figure*}
  \centering
  \includegraphics[width=0.6\textwidth]{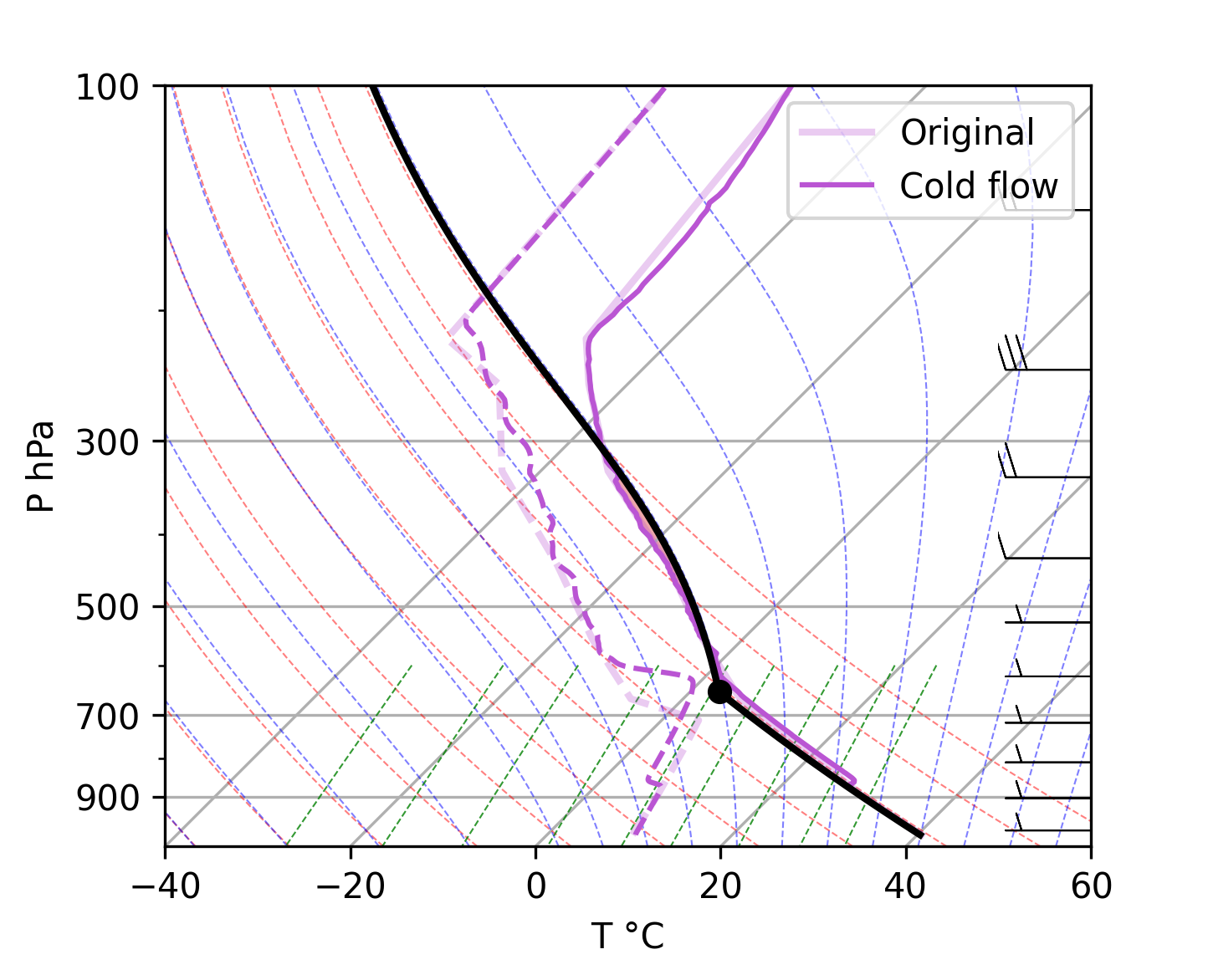} % for an image file named example_figure.*
  % Pick an appriopriate width for the size of the image

  % Captions go below figures
  \caption{\textbf{Thermodynamic profile from the initial and inflow conditions of baseline simulation.} The solid traces are the environmental temperature, the dashed traces are the dew point temperature, and the black trace is the temperature of a parcel adiabatically lifted from the surface.}
  \label{fig:skew_t_baseline} % give each figure a logical label name
\end{figure}

\begin{figure}
  \centering
  \includegraphics[width=0.6\textwidth]{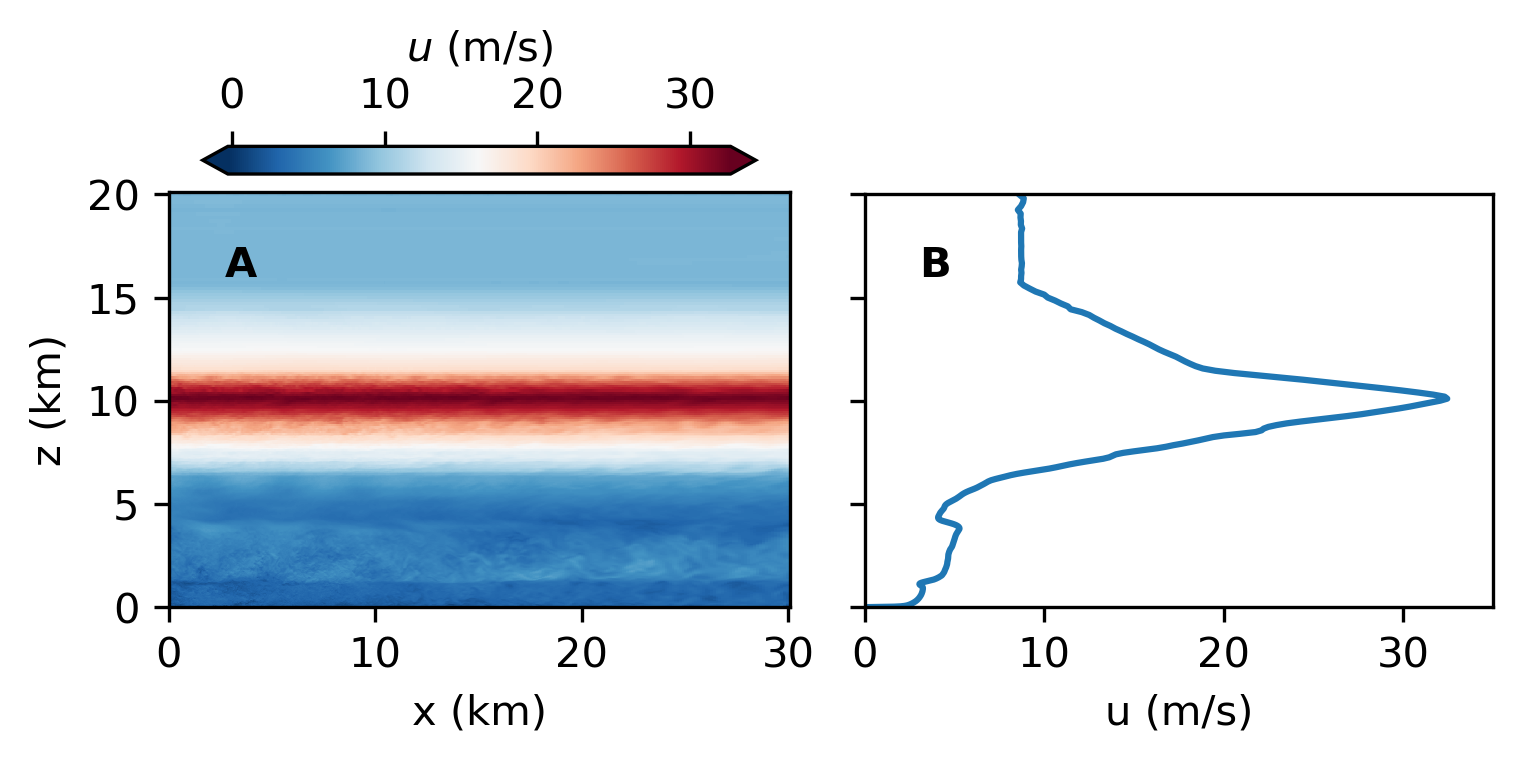}
  \caption{\textbf{The boundary layer profile for the streamwise velocity for the simulation with the original sounding profile.} \textbf{A}, a cross section of the streamwise velocity at the $y=0$ plane. \textbf{B}, the mean of the streamwise velocity across the horizontal ($x-y$) plane.}
  \label{fig:u_init_and_mean}
\end{figure}

Simulations performed in this study are configured based on an approximated condition of the Williams Flat fire~\cite{Peterson2022-wm} on 8 August 2019 using the 00Z 09 Aug 2019 sounding from Spokane, WA. The atmospheric condition of the baseline simulation is shown in fig.~\ref{fig:skew_t_baseline} by the pale red and green curves for the temperature and dew point temperature, respectively. The convective available potential energy (CAPE) for this sounding is $152~\mathrm{J~kg^{-1}}$. The mean wind is assumed to be aligned with the $x$ axis, as shown in fig.~\ref{fig:u_init_and_mean}B.

A precursor simulation initialized with these retrieved conditions is performed in a periodic domain to generate the inflow boundary conditions for the pyroCb simulations. The domain size of the precursor simulation is $10\times 10\times 20\ \text{km}^3$, with a mesh resolution of 5~m in the horizontal directions, and the same mesh as the pyroCb simulations along the vertical dimension. This simulation is conducted for over 30 flow-through times for the turbulent statistics to be fully developed before the inflow profiles are collected. The last snapshot of the precursor simulation is interpolated onto the simulation domain of the pyroCb as the initial condition. The temperature and dew point temperature are shown by the solid red and green curves on fig.~\ref{fig:skew_t_baseline}. The CAPE increases to $198~\mathrm{J~kg^{-1}}$ as the flow field turbulent mixing develops.

A two-layer fuel distribution is employed to model the understory and crown fire~\cite{Dupuy2005-qp}. The fuel density and height are inferred from the Landfire dataset~\cite{US-Department-of-Interior-Geological-Survey-and-US-Department-of-AgricultureUnknown-gy} and are assumed to be distributed homogeneously across the space. A dense layer of fuel with a bulk density $7.96\ \mathrm{kg~m^{-3}}$ is set below 0.5 m. A canopy layer with a bulk fuel density of $0.68\ \mathrm{kg~m^{-3}}$ is set from 0.5 m to 18.8~m~\cite{Peterson2022-wm}.

An inflow-outflow boundary condition is imposed along the $x$ dimension. Turbulent inflow profiles of the 3 velocity components ($u$, $v$, $w$), the liquid-ice potential temperature ($\theta_\text{li}$), and the total humidity ($q_t$), which are collected for 100 minutes of physical time from the precursor simulation, are imposed on the left end of the $x$ axis. To simulate various wind speeds and shears, the mean wind profile is rescaled in the perturbation experiments, while the structure and magnitude of turbulent eddies is preserved~\cite{Wang2024-ix}. A homogeneous Neumann boundary condition is applied on the right end of the $x$ axis. A free-slip adiabatic wall boundary condition is applied at the bottom surface, with a canopy drag force enforced in the fuel layer to model the shear stress produced by the atmospheric boundary layer. A Rayleigh damping layer of 25\% of the domain height is introduced at the top of the domain to eliminate gravity waves. A periodic boundary condition is applied along the $y$ dimension for simplicity.

In all simulations, the ignition is performed by imposing a circular heat source to $\theta_\text{li}$ at 5 km from the inflow with a radius of 800 m, which is comparable to the reported fire area associated with the Region 1 South pyrocumulus cloud \cite{Peterson2022-wm}. The ignition process is conducted following a piecewise linear function in time:
\begin{equation}
    \dot{\omega}_\text{ign}(t) =
    \begin{cases}
      \dot{\omega}_\text{ign,max}\frac{t}{t_0}, & t\in[0, t_0); \\
      \dot{\omega}_\text{ign,max}, & t\in [t_0, t_1);  \\
      \dot{\omega}_\text{ign,max}\frac{t_2 - t}{t_2 - t_1}, & t\in [t_1, t_2],  \\
    \end{cases}
    \label{eq:ignition}
\end{equation}
where $\dot{\omega}_\text{ign,max} = 2.425~\mathrm{kW~m^{-3}}$ is the maximum heat source, $t_0 = 10\text{ s}$,  $t_1 = 610\text{ s}$, and $t_2 = 620\text{ s}$.

\subsubsection*{\label{SUB_SEC_GOVERNING_EQS}Governing equations}

The spatio-temporal evolution of the fire-spread dynamics and the coupling to the atmospheric flow is the described by the LES-filtered conservation equations for mass, momentum, potential temperature, and scalars~\cite{Klemp1978-mm,Linn2005-gm,Pressel2015-gh}:
\begin{eqnarray}
  % \begin{aligned}
   \partial_t \overline{\rho} + \nabla\cdot(\overline{\rho}\widetilde{\boldsymbol{u}})&=& S_F + S_{W,\text{dehy}}  - \sum_{p\in\{r,s\}} \overline{H_p},  \label{eq:mass}\\
   \partial_t {(\overline{\rho} \widetilde{\boldsymbol{u}})} +\nabla\cdot(\overline{\rho}\widetilde{\boldsymbol{u}}\otimes\widetilde{\boldsymbol{u}})&=& -\nabla\overline{p_d} + \nabla\cdot\overline{\tau}+\left[\overline{\rho}-\rho\left(z\right)\right]g\hat{k}_z + \mathbf{f}_\text{D},  \label{eq:momentum}\\
   \partial_t (\overline{\rho}\widetilde{\theta_\text{li}}) + \nabla\cdot(\overline{\rho}\widetilde{\boldsymbol{u}}\widetilde{\theta_\text{li}})&=& \nabla\cdot\overline{\boldsymbol{q}}  \nonumber \\
     &\hspace*{-3cm}+& \hspace*{-1.8cm}\frac{1}{c_{p,m}\Pi}\left[h a_v (T_s - \widetilde{T}) + \dot{q}_\text{rad} + (1 - \Theta) H_f \widetilde{\dot{\omega}} + \left(L_v \overline{H_r} + L_s \overline{H_s}\right)\right]  \label{eq:energy}\\
   \partial_t (\overline{\rho}\widetilde{Y_O}) + \nabla\cdot(\overline{\rho}\widetilde{\boldsymbol{u}}\widetilde{Y_O}) &=& \nabla\cdot\overline{\mathbf{j}_{Y_O}} + \overline{\rho}\widetilde{\dot{\omega}_O},  \label{eq:oxygen}\\
   \partial_t (\overline{\rho}\widetilde{q_t}) + \nabla\cdot(\overline{\rho}\widetilde{\boldsymbol{u}}\widetilde{q_t}) &=& \nabla\cdot\overline{\mathbf{j}_{q_t}} - \sum_{p\in\{r,s\}} \overline{H_p} + S_{W,\text{dehy}} + S_{W,\text{comb}},  \label{eq:q_t}\\
   \partial_t (\overline{\rho}\widetilde{q_p}) + \nabla\cdot\left[\overline{\rho}(\widetilde{\boldsymbol{u}} - w_p\hat{k}_z)\widetilde{q_p}\right] &=& \nabla\cdot\overline{\mathbf{j}_{q_p}} + \overline{H_p},  \label{eq:q_p}
  % \end{aligned}
\end{eqnarray}
where $\rho$ is the density, $\boldsymbol{u}$ is the velocity vector, $\theta_\text{li}$ is the liquid-ice potential temperature, $Y_O$ is the mass fraction of the oxidizer, $q_t$ is the total specific humidity, and $q_p$ is the specific humidity for precipitation with $p\in\{r,s\}$ representing rain and snow, respectively. In these equations, $p_d$ is the hydrodynamic pressure; $\tau$ is the shear stress tensor; $g$ is the  gravitational acceleration, $\hat{k}_z$ is the vertical unit vector; $\mathbf{f}_D=-\overline{\rho}c_d a_v \|\widetilde{\mathbf{u}}\|\widetilde{\mathbf{u}}$ is the canopy drag force with $c_d=0.1$; $\mathbf{q}$ is the diffusive heat flux; $\mathbf{j}_\phi$ is the diffusive flux of scalar $
\phi$; $\widetilde{\dot{\omega}}$ is the source term due to combustion; and $H_p = -E_p + A_p + C_p$ represents all microphysics source term owing to evaporation/sublimation ($E_p$), autoconversion ($A_p$), and accretion ($C_p$). In the potential temperature equation, $h$ is the convection coefficient of air; $a_v$ is the area to v
olume ratio of the fuel element; $T_s$ and $T$ are the temperature for the solid phase and gaseous phase, respectively; $\dot{q}_\text{rad}$ is a heat sink due to radiation; $\Theta = 1 - \rho_f / \rho_{f,0}$ is the fraction of the heat release that contributes to the increase of the solid phase temperature; $H_f$ is the heat of combustion; $L_v$ and $L_s$ are the latent heat of vaporization and sublimation, respectively; and $\Pi = (p(z) / p_0)^{R_m / c_{p, m}}$ is the Exner function. In the mass and $q_t$ transport equations, $S_F$, $S_{W, \text{comb}}$, and $S_{W, \text{dehy}}$ are the rates of solid-fuel consumption, combustion water production, and fuel-moisture dehydration, respectively.

In the LES formulation, the Favre-filtered value for a generic variable $\phi$ is defined as $\widetilde{\phi}=\overline{\rho\phi}/\overline{\rho}$, where the overbar denotes the Reynolds filtering. The shear stress and diffusive fluxes are represented with the Boussinesq approximation, which are computed as:
\begin{eqnarray}
% \begin{align}
    \overline{\tau} &=& 2\overline{\rho}(\nu + \nu_t)\widetilde{S},  \\
    \overline{\mathbf{q}} &=& \overline{\rho}(\lambda + \lambda_t)\nabla\widetilde{\theta},  \\
    \overline{\mathbf{j}_\phi} &=& \overline{\rho}(\alpha + \alpha_t)\nabla\widetilde{\phi},
% \end{align}
\end{eqnarray}
where $\widetilde{S} = [\nabla\widetilde{\mathbf{u}} + (\nabla\widetilde{\mathbf{u}})^T]/2 + (\nabla\cdot\widetilde{\mathbf{u}} I) / 3$ is the strain rate tensor, and $\mu$, $\lambda$, and $\alpha_\phi$ are the dynamic viscosity, thermal conductivity, and diffusivity for a generic scalar $\phi$, respectively. The kinematic eddy viscosity is computed using the Smagorinsky model as $\nu_t=(C_s \Delta)^2|\widetilde{S}|$ with $C_s = 0.18$. The turbulent conductivity and diffusivity are derived from the eddy viscosity with constant turbulent Prandtl and Schmidt numbers, which are $\lambda_t = \nu_t / \text{Pr}_t$ and $\alpha_t = \nu_t / \text{Sc}_t$, respectively.

In this study, eqs.~\ref{eq:mass} to~\ref{eq:q_p} are solved with a low-Mach number approximation. The pressure is decomposed into hydrostatic and hydrodynamic components, $p=p(z)+p_d$, with the hydrostatic pressure computed as $p(z)=p_0[1-gz/(c_{p,m}\theta_\infty)]^{1/\kappa}$, where $\theta_\infty$ is the potential temperature of the ambient air that is obtained from the initial sounding, $p_0$ is the atmospheric pressure on the ground level, and $\kappa=R_m/c_{p,m}$ is the ratio between the gas constant and specific heat of the mixture. The temperature ($T$) and potential temperature ($\theta$) are related to the liquid-ice potential temperature ($\theta_\text{li}$) as:
\begin{eqnarray}
% \begin{align}
    \theta_\text{li} &=& \frac{T}{\Pi} \left(1 - \frac{L_{v,0} q_l + L_{s,0} q_i}{c_{p,m}T} \right),  \\
    T &=& \theta\Pi.
% \end{align}
\end{eqnarray}

\subsubsection*{The microphysics model}

We use the one-moment microphysics model from~\cite{CloudMicrophysics25a}, which derives from~\cite{Kessler1995-xx} and~\cite{Grabowski98a}. For simplicity, the overbars and tildes on variables are omitted in the following equations. In this model, the particle mass ($m$), cross-section area ($a$), and sedimentation velocity ($v$) are represented as functions of the particle radius~\cite{Grabowski98a},
\begin{equation}
  \phi_\eta = \phi_{0,\eta} \left(\frac{r}{r_{0,\eta}}\right)^{\phi_{e,\eta}},
\end{equation}
where $\phi\in\{m, a, v\}$ and $\eta\in\{r,s,i\}$ represent rain, snow, and ice, respectively, and $\phi_{0,\eta}$ and $\phi_{e,\eta}$ are constant coefficients. The slope parameter of the particle size distribution function is computed as:
\begin{equation}
    \lambda_\eta = \left[\frac{m_{0,\eta} n_{0,\eta} \Gamma(m_{e,\eta} + 1)}{\rho q_\eta} \right]^{1/(m_{e,\eta}+1)}.
\end{equation}
With this, the particle size distributions are assumed to follow the Marshall-Palmer distribution:
\begin{equation}
    n_\eta = n_{0,\eta} \exp(-\lambda_\eta r).
\end{equation}

For $p\in\{r,s\}$ (rain and snow), the mass-weighted terminal velocity is computed as:
\begin{equation}
    w_p = v_{0,p} \frac{\Gamma(m_{e,p} + v_{e,p} + 1)}{\Gamma(m_{e,p} + 1) \lambda_p^{-v_{e,p}}}.
\end{equation}
The rates of autoconversion, accretion for converting rain/snow to cloud droplets/ice, and  evaporation/sublimation are computed~\cite{Kessler1995-xx,Grabowski98a,Kaul2015-ky,CloudMicrophysics25a} as
\begin{eqnarray}
    A_p &=& \frac{\max(q_c - q_{c,\text{threshold}}, 0)}{\tau_{c,\text{aut}}},  \\
    C_p &=& \frac{\pi}{4}n_{0,p}v_{0,p}q_{c}E_{cp}\Gamma(v_e + 3)\lambda_{p}^{-(v_e + 3)},  \\
    E_p &=& \frac{4\pi n_{0,p}}{\rho} (S - 1) G_p F_p \lambda_p^{-2},
\end{eqnarray}
with $c=l$ when $p=r$, and $c=i$ when $p=s$. Here, $S=q_v/q_{v,sat}$ is the saturation ratio, and $G_p$ is a thermodynamic quantity that measures the combined effect of thermal conductivity and water diffusivity:
\begin{equation}
    G_p = \left[\frac{L_h}{K_a T} \left(\frac{L_h}{R_v T} - 1\right) + \frac{R_v T}{p_{v,sat} \alpha_v}\right]^{-1}.
\end{equation}
Here, the specific latent heat $L_h=L_v$ when $p=r$ and $L_h=L_s$ when $p=s$; $K_a$ is the thermal conductivity of air; $R_v$ is the gas constant of water vapor; and $\alpha_v$ is the thermal diffusivity of water vapor. The ventilation factor is computed as
\begin{equation}
    F_p = a_{\text{vent},p} + b_{\text{vent},p} \left(\frac{\nu}{\alpha_v} \right)^{1/3} \left(\frac{2 v_{0,p}}{\nu\lambda_p} \right)^{1/2} \left(\frac{1}{r_{0,p}\lambda_p} \right)^{v_e/2} \Gamma\left(\frac{v_{e,p} + 5}{2}\right),
\end{equation}
where $\nu$ is the kinematic viscosity of the air.

\subsubsection*{Thermodynamics}
The temperature ($T$) and the partitioning of total water fraction into liquid ($q_l$) and ice ($q_i$) phases are obtained from $\theta_\text{li}$, $q_t$, and $p(z)$ through saturation adjustment~\cite{Tao1989-ov}, where we assume that the liquid-ice potential temperature is at saturation when condensate is present, i.e.,
\begin{equation}
    \theta_\text{li} - \theta_\text{li}^* = 0,
    \label{eq:sat_adj}
\end{equation}
with $\theta_\text{li}^* = \theta_\text{li}(T, q_l^*, q_i^*, p(z))$ and
\begin{eqnarray}
% \begin{align}
    q_l^* &=& \max [q_t - q_v^*(T, p(z))] \mathcal{H}(T - T_f),  \\
    q_i^* &=& \max [q_t - q_v^*(T, p(z))] \mathcal{H}(T_f - T),
% \end{align}
\end{eqnarray}
where $q_v^*$ is the saturation specific humidity, $\mathcal{H}$ is the Heaviside function, and $T_f$ is the freezing point temperature. Assuming that the gases are calorically perfect, $q_v^*$ is obtained from integrating the Clausius-Clapeyron equation. Equation~\ref{eq:sat_adj} is solved numerically with secant method. With this, the density is obtained from the ideal gas law as:
\begin{equation}
    \rho = \frac{p(z)}{R T}.
\end{equation}

\subsubsection*{Physics-based combustion model}
The combustion is described by a one-step global reaction that represents both the solid-fuel pyrolysis and gas-fuel reaction~\cite{Linn1997-av}:
\begin{equation}
    \nu_F F + \nu_O O \rightarrow \nu_P P,
    \label{eq:chemical_reaction}
\end{equation}
where $\nu_F$, $\nu_O$, and $\nu_P$ are the stoichiometric coefficients of the fuel $F$ in the solid phase, the oxygen $O$, and the combustion product $P$ in the gas phase, respectively. The reaction rate of the combined process of pyrolysis and gas-phase combustion is modeled as
\begin{equation}
    \widetilde{\dot{\omega}} = \frac{c_F \rho_F\overline{\rho}\widetilde{Y_O}\nu_t \Psi_S \lambda_{OF}}{\rho_\infty s_x^2},
    \label{eq:reaction_rate}
\end{equation}
where $c_F = 0.5$ is an empirical scaling coefficient, $\rho_F$ is the bulk fuel density that is defined as the ratio between the fuel load and the height of the fuel, $\overline{\rho}$ is the filtered gas phase density, $\rho_\infty = 1~\mathrm{kg~m^{-3}}$ is the reference density, $\nu_t$ is the turbulent viscosity, and $s_x = 0.05\ \text{m}$ is an empirical coefficient to parameterize the characteristic turbulence scale. In this model, $\lambda_\text{OF}$ is introduced as a coefficient that maximizes the reaction rate, which is formulated as
\begin{equation}
    \lambda_\text{OF} = \frac{\rho_F\overline{\rho}\widetilde{Y_O}}{(\rho_F / \nu_F + \overline{\rho}\widetilde{Y_O}/\nu_O)^2}.
\end{equation}
The variable $\Psi_S$ in eq.~\ref{eq:reaction_rate} represents the ignited volume fraction, which is modeled as a piecewise-linear function of the solid-phase tempearture $T_s$:
\begin{equation}
    \Psi_S = \max\left( \min \left( \frac{T_s - T_{\min}}{T_{\max} - T_{\min}}, 1\right), 0 \right),
\end{equation}
where $T_{\min} = 323\ \text{K}$ and $T_{\max} = 700\ \text{K}$. The rate of fuel consumption is then computed as
\begin{equation}
    S_F = \nu_F \widetilde{\dot{\omega}}.
\end{equation}

Note that eq.~\ref{eq:chemical_reaction} can be approximated as the chemical reaction as follows~\cite{Cunningham2009}:
\begin{equation}
    \text{C}_6 \text{H}_{10} \text{O}_5 + 6\text{O}_2 = 6\text{CO}_2 + 5\text{H}_2\text{O}.
    \label{eq:chemical_reaction_approx}
\end{equation}
Based on eq.~\ref{eq:chemical_reaction_approx}, the combustion water is estimated from the total combustion product that is obtained from the simulation, which is
\begin{equation}
    S_{W,\text{comb}} = \frac{5 W_{\text{H}_2\text{O}}}{6 W_{\text{CO}_2} + 5 W_{\text{H}_2\text{O}}} \widetilde{\dot{\omega}_p},
\end{equation}
where $W_{\text{H}_2\text{O}}$ and $W_{\text{CO}_2}$ are the molecular mass of water and carbon dioxide, respectively, and $\widetilde{\dot{\omega}_p}=(\nu_F + \nu_O)\widetilde{\dot{\omega}}$ is the rate of formation of the combustion product.

The fuel moisture is modelled as the bulk density of liquid water, $\rho_W$. The rate of vaporization is computed as
\begin{equation}
  S_{W,\text{dehy}} = \rho_{W} \frac{A_{\text{deh}}}{\sqrt{T_s}}\exp\left\{-\frac{T_{\text{deh}}}{T_s}\right\}\;,
  \label{eq:deh_rate}
\end{equation}
with $A_{\text{deh}} = 6.05\times10^5~\mathrm{K^{1/2}~s^{-1}}$ and $T_{\text{deh}}=5956$\,K.

The solid states, including the bulk fuel and moisture densities, and the temperature of the fuel, are modelled with the following ordinary differential equations~\cite{Linn2005-gm}:
\begin{eqnarray}
% \begin{align}
    d_t \rho_F &=& -S_F,  \label{eq:rho_f} \\
    d_t \rho_W &=& -S_{W,\text{dehy}},  \label{eq:rho_w} \\
    (c_{p,F}\rho_F + c_{p,W}\rho_W) d_t T_s &=& \dot{q}_\text{rad,s} + h a_v (\widetilde{T} - T_s) \nonumber \\
    &\hspace*{-1cm}-&\hspace*{-0.8cm} (L_v + c_{p,W} T_\text{vap})S_{W,\text{dehy}} + (\Theta H_f - c_{p,F} T_\text{pyr}\nu_F)\widetilde{\dot{\omega}},  \label{eq:t_s}
% \end{align}
\end{eqnarray}
where $c_{p,F}$ and $c_{p,W}$ are the specific heat of the fuel and liquid water, respectively, $L_v$ is the latent heat of vaporization, $T_\text{vap}$ is the temperature of vaporization, and $T_\text{pyr}$ is the temperature at which the solid fuel begins to pyrolyse.

The radiation sink terms in eqs.~\ref{eq:energy} and~\ref{eq:t_s} are modelled by a grey-gas model as $\dot{q}_\text{rad}(T) = -(\sigma k/\zeta)(T^4 - T_\infty^4)$, with $\sigma$ being the Boltzmann constant, $k$ being a coefficient that models the turbulence-radiation interaction and is set to 1 for a balanced interaction~\cite{Linn1997-av}; $T_\infty$ is the ambient temperature; and $\zeta$ is a characteristic length scale of the fuel elements that is set to 0.5 m.

The source terms due to combustion and vaporization are incorporated in the gas phase equations through a Lie splitting scheme~\cite{Trotter1959-ln}, where the hydrodynamics are integrated following the source terms.

%%%%%%%%%%%%%%%% SUPPLEMENTARY TEXT %%%%%%%%%%%%%%%

\subsection*{Supplementary Text}

\subsubsection*{Lagrangian Parcel Tracking}
To quantify the dynamics of the fire-atmosphere interactions, we performed a detailed Lagrangian parcel tracking study. For this analysis, $10^5$ passive tracer parcels were uniformly seeded within a horizontal domain encompassing the final fire scar. Parcels were released every minute after the initial 20 minutes of simulation time and their trajectories were traced for 30 minutes, a duration sufficient to capture the key stages of pyroCb development. Along these trajectories, we collected detailed energy and humidity budget information (eqs.~\ref{eq:total_energy_density_parcel_with_solid} and~\ref{eq:q_t_parcel}). 

\subsubsection*{\label{SUP_SEC_PARCEL}Parcel Analyses}
\begin{figure}
    \centering
    \includegraphics[width=0.6\linewidth]{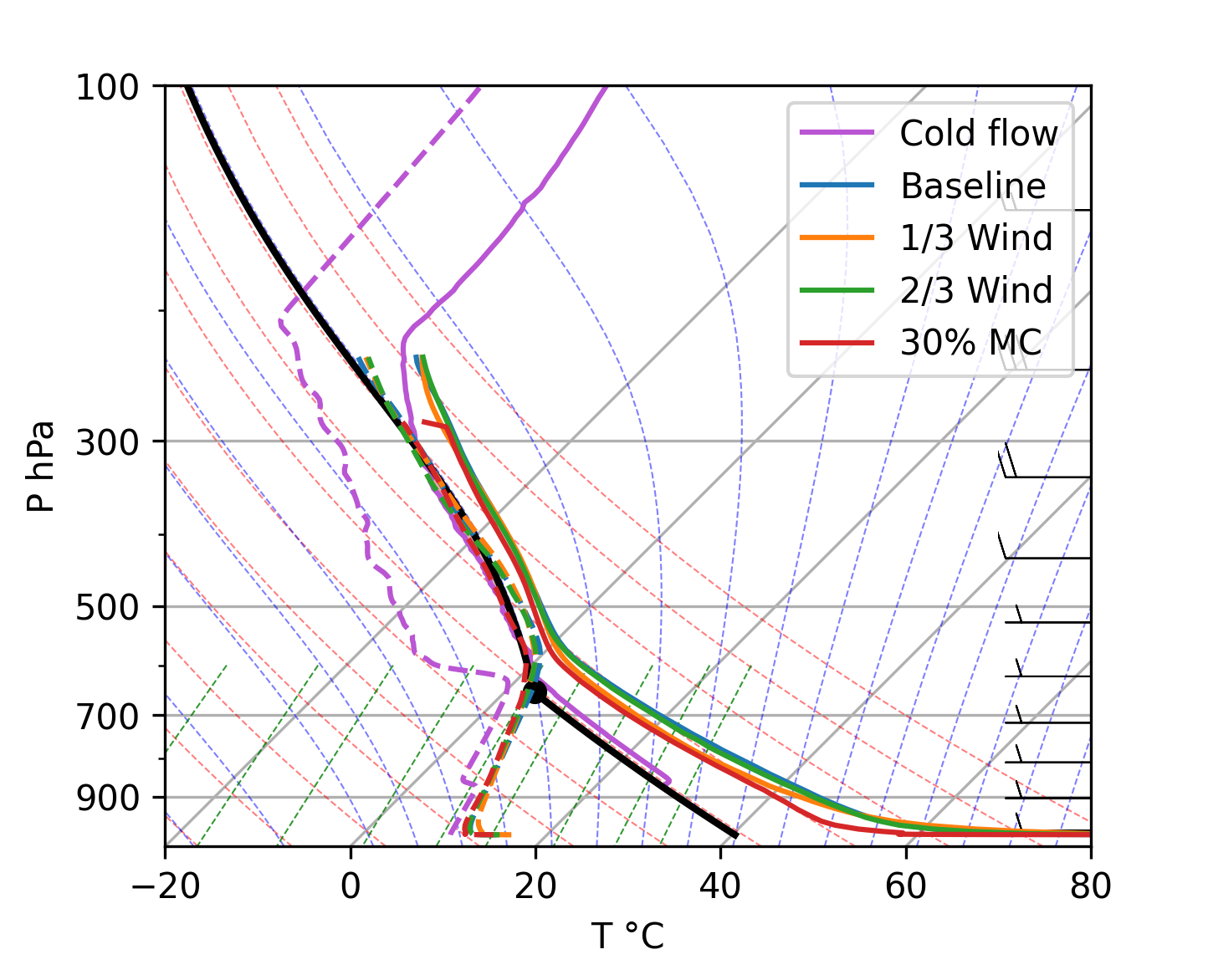}
    \caption{\textbf{The thermodynamic plot of the mean parcels.} The solid traces are the environmental temperature, the dashed trace are the dew point temperature, and the black trace is the temperature of a parcel adiabatically lifted from the surface.}
    \label{fig:skew_t_parcel}
\end{figure}

\begin{figure}
  \centering
  \includegraphics[width=0.9\textwidth]{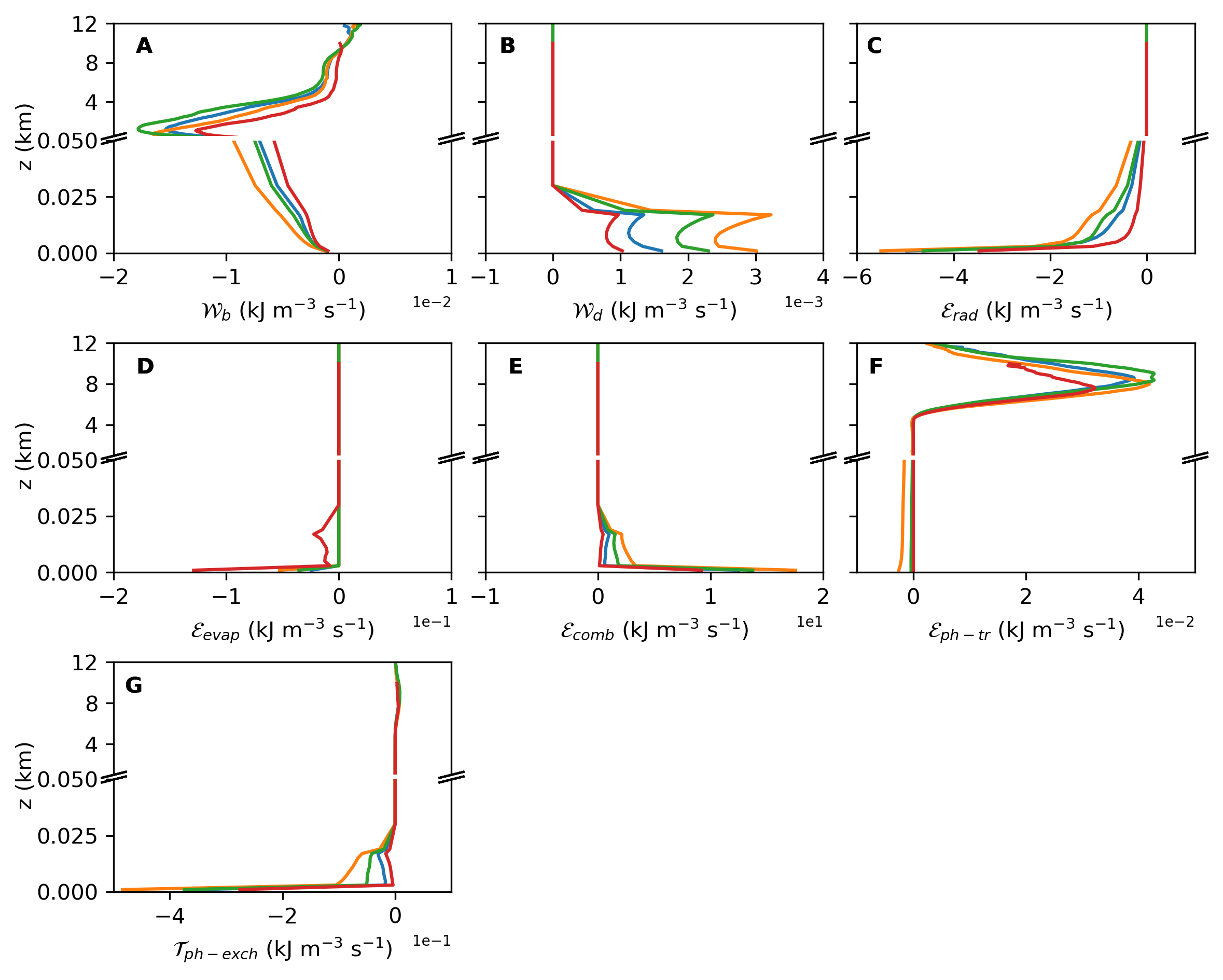}
  \caption{\textbf{The energy budget following eq.~\ref{eq:total_energy_density_parcel_with_solid}.} Key terms influencing the attenuation mechanism are shown: \textbf{A}, Buoyancy work ($\mathcal{W}_b$), the primary driver of convection. \textbf{B}, Canopy-drag heating ($\mathcal{W}_d$). \textbf{C}, Radiative heat flux ($\mathcal{E}_\text{rad}$). \textbf{D}, Energy consumed by fuel-moisture vaporization ($\mathcal{E}_\text{evap}$), which acts as a dominant heat sink in the 30\% MC case. \textbf{E}, Heat release from combustion ($\mathcal{E}_\text{comb}$). \textbf{F}, Latent heat release from precipitation condensation ($\mathcal{T}_\text{ph-tr}$). \textbf{G}, Energy transfer due to mass exchange ($\mathcal{T}_\text{ph-exch}$). Legends used by the lines for the four simulation cases are blue: baseline; orange: 1/3 wind; green: 2/3 wind; red: 30\% MC.}
  \label{fig:energy_budget_ext}
\end{figure}

\begin{figure}[!htb!]
  \centering
  \includegraphics[width=0.9\textwidth]{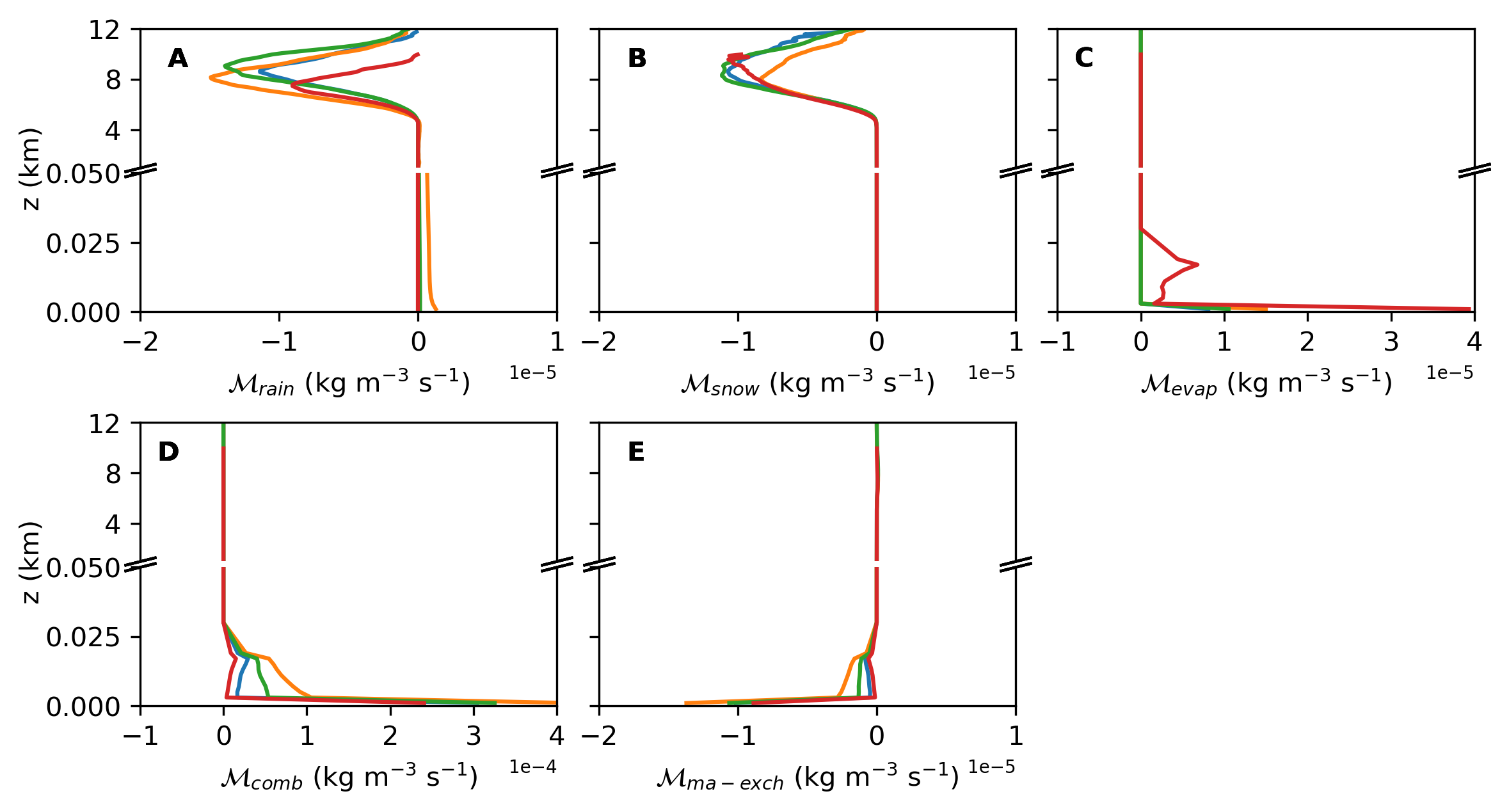}
  \caption{\textbf{The humidity budget following eq.~\ref{eq:q_t_parcel}}. Key terms influencing the attenuation mechanism are shown: \textbf{A}, Rain ($\mathcal{M}_\text{rain}$). \textbf{B}, Snow ($\mathcal{M}_\text{snow}$). \textbf{C}, Vaporization of fuel moisture ($\mathcal{M}_\text{evap}$). \textbf{D}, Combustion product ($\mathcal{M}_\text{comb}$). \textbf{E}, Moisture transfer due to mass exchange ($\mathcal{M}_\text{ma-exch}$). Legends used by the lines for the four simulation cases are blue: baseline; orange: 1/3 wind; green: 2/3 wind; red: 30\% MC.}
  \label{fig:water_budget_ext}
\end{figure}

A budget analysis was performed along Lagrangian parcel trajectories to quantify the contributions of different physical processes to the changes in energy and total humidity. The horizontal mean of the energy (eq.~\ref{eq:total_energy_density_parcel_with_solid}) and humidity (eq.~\ref{eq:q_t_parcel}) budget terms are computed statistically from all parcel trajectories, and are plotted as a function of height in figs.~\ref{fig:energy_budget_ext} and~\ref{fig:water_budget_ext}, respectively. The total budget terms are therefore computed by integrating the mean budget profiles along the vertical direction ($z$), and are shown in Figs.~\ref{fig:energy_budget}B and C. The mean thermodynamic properties of the updraft parcels for each simulation case are plotted on a skew-T diagram (fig.~\ref{fig:skew_t_parcel}), with key variables summarized in table~\ref{tab:skew_t_parcel}.

\paragraph{Energy budget equation}

The budget analysis for total energy is derived from the transport equation for liquid-ice potential temperature, $\theta_\text{li}$ (eq.~\ref{eq:energy}). Multiplying by $c_{p,m}$ on both sides of eq.~\ref{eq:energy}, and assuming that the gas constant and heat capacity for the mixture are equal to those of the dry air ($R_m=R_d$ and $c_{p,m}=c_{p,d}$), we have
\begin{equation}
    c_{p,m} \partial_t ({\rho \theta_\text{li}}) = \partial_t \left\{{\rho \left[c_{p,m} \theta - \frac{1}{\Pi}\left(L_{v,0} q_l + L_{s,0} q_i\right)\right]}\right\} ,
\end{equation}
and
\begin{eqnarray}
    &&c_{p,m}\nabla\cdot\left( \rho \boldsymbol{u} \theta_\text{li}\right) = \nabla\cdot\left\{\rho \boldsymbol{u}\left[c_{p,m}\theta - \frac{1}{\Pi}\left(L_{v,0} q_l + L_{s,0} q_i\right)\right]\right\}  \nonumber \\
    &=&\nabla\cdot\left(\rho \boldsymbol{u}c_{p,m}\theta\right)  - \frac{1}{\Pi}\nabla\cdot\rho\boldsymbol{u}\left(L_{v,0} q_l + L_{s,0} q_i\right) - \rho\boldsymbol{u}\left(L_{v,0} q_l + L_{s,0} q_i\right)\cdot\nabla\frac{1}{\Pi}.
    \label{eq:energy_deriv_conv}
\end{eqnarray}
Note that, because the basic-state pressure in the Exner function is hydrostatically balanced ($p$ is a function of $z$ only),
\begin{eqnarray}
% \begin{align}
    \nabla\frac{1}{\Pi} &=& -\frac{1}{\Pi^2}\frac{R_m}{c_{p,m}} \frac{1}{p_0} \left(\frac{p}{p_0}\right)^{R_m/c_{p,m} - 1} \frac{\partial p}{\partial z} \hat{k}_z  \nonumber \\
    &=&-\frac{1}{\Pi^2}\frac{R_m}{c_{p,m}} \frac{1}{p} \left(\frac{p}{p_0}\right)^{R_m/c_{p,m}} \frac{\partial p}{\partial z} \hat{k}_z  \nonumber \\
    &=&-\frac{1}{\Pi^2}\frac{R_m}{c_{p,m}} \frac{1}{\rho(z) R_m T(z)} \Pi \left( -\rho(z) g\right)\hat{k}_z  \nonumber \\
    &=&\frac{1}{\Pi}\frac{1}{c_{p,m}T(z)}g\hat{k}_z,
% \end{align}
\end{eqnarray}
eq.~\ref{eq:energy_deriv_conv} can be simplified as
\begin{eqnarray}
% \begin{align}
    c_{p,m}\nabla\cdot\left( \rho \boldsymbol{u} \theta_\text{li}\right) &=&\nabla\cdot\rho \boldsymbol{u}\left(c_{p,m}\theta\right)  - \frac{1}{\Pi}\nabla\cdot\rho\boldsymbol{u}\left(L_{v,0} q_l + L_{s,0} q_i\right)  \nonumber \\
    &\hspace*{-1cm}-&\hspace*{-0.8cm} \rho\boldsymbol{u}\left(L_{v,0} q_l + L_{s,0} q_i\right)\cdot\frac{1}{\Pi}\frac{1}{c_{p,m}T(z)}g\hat{k}_z  \nonumber \\
    &\hspace*{-1cm}=&\hspace*{-0.8cm} \nabla\cdot\rho \boldsymbol{u}\left(c_{p,m}\theta\right) - \frac{1}{\Pi}\sum_{c\in\{l,i\}}L_{f,0}\left[\nabla\cdot\rho\boldsymbol{u}q_c + \frac{q_c}{c_{p,m}T(z)}\rho\boldsymbol{u}\cdot g\hat{k}_z  \right],
    \label{eq:energy_conv}
% \end{align}
\end{eqnarray}
where $c\in\{l,i\}$ is the cloud fraction in liquid and ice phases, respectively; and 
\begin{equation}
    L_{f,0} =
    \begin{cases}
        L_{v,0}, &c=l; \\
        L_{s,0}, &c=i,
    \end{cases}
\end{equation}
is the latent heat of vaporization and fusion correspondingly. By defining the mass-specific energy as $e=c_{p,m}\theta$~\cite{Jacobson1999-xq} (where $\theta$ is potential temperature), eq.~\ref{eq:energy} then becomes
\begin{eqnarray}
% \begin{align}
  \partial_t \overline{\rho}\widetilde{e} &+& \nabla\cdot(\overline{\rho}\widetilde{\boldsymbol{u}}\widetilde{e}) = \nabla\cdot\left[\overline{\rho} \left(\lambda + \lambda_t\right)\nabla \widetilde{e}\right]  \nonumber \\
  &+& \frac{1}{\Pi}\left[h a_v (T_s - \widetilde{T}) + \dot{q}_\text{rad} + (1 - \Theta) H_f \widetilde{\dot{\omega}} + \left(L_v \overline{H_r} + L_s \overline{H_s}\right)\right] \nonumber \\
  &+&\sum_{c\in\{l,i\}}L_{f,0}\left\{ \frac{1}{\Pi} \left[ \partial_t \overline{\rho}\widetilde{q_c} + \nabla\cdot\left(\overline{\rho}\widetilde{\boldsymbol{u}}\widetilde{q_c}\right) + \left(\frac{ \widetilde{q_c}}{c_{p,m}T(z)}\right)\overline{\rho}\widetilde{\boldsymbol{u}}\cdot g\hat{k}_z\right] \right. \nonumber \\
  &&\left. -\nabla\cdot\left[\overline{\rho} \left(\lambda + \lambda_t\right)\nabla \left(\frac{\widetilde{q_c}}{\Pi}\right)\right]\right\},
  \label{eq:energy_density}
% \end{align}
\end{eqnarray}

Adding the kinetic energy equation, which is obtained by applying a dot product between the velocity and eq.~\ref{eq:momentum}, to eq.~\ref{eq:energy_density} provides an equation for the total energy density $e_t = c_{p,m}\theta + \frac{1}{2}\boldsymbol{u}^2$:
\begin{eqnarray}
% \begin{align}
  \partial_t \left(\overline{\rho}\widetilde{e_t}\right) &+& \nabla\cdot(\overline{\rho}\widetilde{\boldsymbol{u}}\widetilde{e_t}) = \nabla\cdot\left[\overline{\rho} \left(\lambda + \lambda_t\right)\nabla \widetilde{e}\right]  \nonumber \\
  &+& \left[-\boldsymbol{u}\cdot\nabla\overline{p_d} + \boldsymbol{u}\cdot\nabla\overline{\tau} + \left(\rho - 
 \rho(z)\right)\widetilde{\boldsymbol{u}}\cdot g\hat{k}_z + \boldsymbol{u}\cdot\boldsymbol{\mathrm{f}_D}\right]\nonumber \\
  &+& \frac{1}{\Pi}\left[h a_v (T_s - \widetilde{T}) + \dot{q}_\text{rad} + (1 - \Theta) H_f \widetilde{\dot{\omega}} + \left(L_v \overline{H_r} + L_s \overline{H_s}\right)\right] \nonumber \\
  &+&\sum_{c\in\{l,i\}}L_{f,0}\left\{ \frac{1}{\Pi} \left[ \partial_t \overline{\rho}\widetilde{q_c} + \nabla\cdot\left(\overline{\rho}\widetilde{\boldsymbol{u}}\widetilde{q_c}\right) + \left(\frac{ \widetilde{q_c}}{c_{p,m}T(z)}\right)\overline{\rho}\widetilde{\boldsymbol{u}}\cdot g\hat{k}_z\right] \right. \nonumber \\
  &&\left.-\nabla\cdot\left[\overline{\rho} \left(\lambda + \lambda_t\right)\nabla \left(\frac{\widetilde{q_c}}{\Pi}\right)\right]\right\}.
  \label{eq:total_energy_density}
% \end{align}
\end{eqnarray}
Terms in braces $\{\cdot\}$ in eq.~\ref{eq:total_energy_density_parcel}, which will be denoted as $\mathcal{R}_{q_c}$, correspond to the energy exchange due to the transport and phase transition of  condensed water in the atmosphere, which we omit in the energy budget study.

Equation~\ref{eq:total_energy_density} can also be expressed as the material derivative. Substituting eq.~\ref{eq:mass} in eq.~\ref{eq:total_energy_density}, we have
\begin{eqnarray}
% \begin{align}
  \overline{\rho}\mathcal{D}_t \widetilde{e_t} &=& \nabla\cdot\left[\overline{\rho} \left(\lambda + \lambda_t\right)\nabla \widetilde{e}\right]  \nonumber \\
  &+& \left[-\boldsymbol{u}\cdot\nabla\overline{p_d} + \boldsymbol{u}\cdot\nabla\overline{\tau} + \left(\rho - 
 \rho(z)\right)\widetilde{\boldsymbol{u}}\cdot g\hat{k}_z + \boldsymbol{u}\cdot\boldsymbol{\mathrm{f}_D}\right]\nonumber \\
  &+& \frac{1}{\Pi}\left[h a_v (T_s - \widetilde{T}) + \dot{q}_\text{rad} + (1 - \Theta) H_f \widetilde{\dot{\omega}} + \left(L_v \overline{H_r} + L_s \overline{H_s}\right)\right] \nonumber \\
  &-& \widetilde{e_t}\left( S_F + S_{W,\text{dehy}} - \sum_{p\in\{r,s\}}\overline{H_p}\right)  \nonumber \\
  &+&\mathcal{R}_{q_c},
  \label{eq:total_energy_density_parcel}
% \end{align}
\end{eqnarray}

To account for the energy associated with the solid fuel and its moisture content, the solid-phase energy equation (eq.~\ref{eq:t_s}) is included. The final form of the total energy budget along a parcel trajectory (expressed as a material derivative, $\mathcal{D}_t$) is:
\begin{eqnarray}
% \begin{align}
  &&\underbrace{\overline{\rho}\mathcal{D}_t \widetilde{e_t} + \rho_F d_t e_F + \rho_W d_t e_W}_{{\cal{E}}_\text{net}: \text{Net Rate of Total Energy Change}}  = \underbrace{\nabla\cdot\left[\overline{\rho} \left(\lambda + \lambda_t\right)\nabla \widetilde{e}\right]}_{{\cal{E}}_\text{diff}: \text{Diffusive Energy Flux}}  \nonumber \\
  &&+ \left[\underbrace{-\boldsymbol{u}\cdot\nabla\overline{p_d}}_{{\cal{W}}_p: \text{Pressure Work}} + \underbrace{\boldsymbol{u}\cdot\nabla\overline{\tau}}_{{\cal{W}}_v: \text{Viscous Dissipation}} + \underbrace{\left(\rho - 
 \rho(z)\right)\widetilde{\boldsymbol{u}}\cdot g\hat{k}_z}_{{\cal{W}}_b: \text{Buoyancy Work}} + \underbrace{\boldsymbol{u}\cdot\boldsymbol{\mathrm{f}_D}}_{{\cal{W}}_d: \text{Drag Force Work}}\right]\nonumber \\
  &&+ \frac{1}{\Pi}\left[\underbrace{\dot{q}_\text{rad,s} + \dot{q}_\text{rad}}_{{\cal{E}}_\text{rad}: \text{Radiation}} - \underbrace{(L_v + c_{p,W} T_\text{vap})S_{W,\text{dehy}}}_{{\cal{E}}_\text{evap}: \text{Fuel Moisture Evaporation Sink}}  \right. \nonumber \\
  &&\left.+\underbrace{(H_f - c_{p,F} T_\text{pyr}\nu_F)\widetilde{\dot{\omega}}}_{{\cal{E}}_\text{comb}: \text{Combustion Heat Release}} + 
  \underbrace{\left(L_v \overline{H_r} + L_s \overline{H_s}\right)}_{{\cal{E}}_\text{ph-exch}: \text{Energy Exchange from Microphysics}}\right] \nonumber \\
  &&- \underbrace{\widetilde{e_t}\left( S_F + S_{W,\text{dehy}} - \sum_{p\in\{r,s\}}\overline{H_p}\right)}_{{\cal{T}}_\text{ma-exch}: \text{Energy Exchange from Mass Transfer}}  \nonumber \\
  &&+\mathcal{R}_{q_c},
  \label{eq:total_energy_density_parcel_with_solid}
% \end{align}
\end{eqnarray}
where $e_F = c_{p,F}\theta_s$ and $e_W = c_{p,W}\theta_s$ are the energies associated with the fuel and moisture, respectively, and $\theta_s = T_s / \Pi$ is the potential temperature of the solid.

The terms in eq.~\ref{eq:total_energy_density_parcel_with_solid} allow for the direct quantification of energy sources (e.g., buoyancy work, heat from combustion) and sinks (e.g., energy consumed by fuel moisture evaporation, radiative cooling). The full set of terms is visualized as vertical profiles in Supplementary fig.~\ref{fig:energy_budget_ext} and as vertical integral in Fig.~\ref{fig:energy_budget}.

\paragraph{Humidity budget equation}

A corresponding budget equation for the total specific humidity, $\widetilde{q_t}$, was derived from the continuity and humidity transport equations (eqs.~\ref{eq:mass} and~\ref{eq:q_t}) to trace the origin of water mass within the pyroCb system:
\begin{eqnarray}
% \begin{align}
  \underbrace{\overline{\rho}\mathcal{D}_t \widetilde{q_t}}_{{{\cal{Q}}_{net}}: \text{Rate of Total Humidity Change}} &=&\underbrace{\nabla\cdot\overline{\mathbf{j}_{q_t}}}_{{{\cal{Q}}_\text{diff}: \text{{Diffusive Flux}}}} - \underbrace{\sum_{p\in\{r,s\}} \overline{H_p}}_{{\cal{M}}_\text{ph-exch}: \text{Microphysics Sink/Source}} \nonumber\\
  &\hspace*{-3cm} +&\hspace*{-2.8cm} \underbrace{S_{W,\text{dehy}}}_{{\cal{M}}_\text{evap}: \text{Microphysics Sink/Source}} + \underbrace{S_{W,\text{comb}}}_{{\cal{M}}_\text{evap}: \text{ Combustion Water Source}} \nonumber\\
 &\hspace*{-3cm}-&\hspace*{-2cm} \underbrace{\widetilde{q_t}\left( S_F + S_{W,\text{dehy}} - \sum_{p\in\{r,s\}}\overline{H_p}\right)}_{{\cal{M}}_\text{ma-exch}: \text{Humidity Exchange from Mass Transfer}},
  \label{eq:q_t_parcel}
% \end{align}
\end{eqnarray}
This equation enables the precise accounting of water vapor sourced from fuel moisture dehydration ($S_{W,\text{dehy}}$) and combustion products ($S_{W,\text{comb}}$), contrasted with the entrainment of ambient environmental humidity, which is implicitly captured by the material derivative on the left-hand side.

\paragraph{Scaling Analysis}

To validate the detailed budget analysis, we performed an order-of-magnitude analysis of the fuel and moisture contributions. Assuming complete combustion and evaporation as the fire front passes, the rate of fuel consumption and moisture evaporation can be approximated as
\begin{eqnarray}
% \begin{align}
    S_{f,\text{comb}} &=& \rho_f \dot{A} h,  \label{eq:fuel_consumption_rate}\\
    S_{w,\text{evap}} &=& \rho_m \dot{A} h,  \label{eq:moisture_vaporization_rate}
% \end{align}
\end{eqnarray}
where $\rho_f$ is the fuel density, $\rho_m$ is the moisture density, $\dot{A}$ is the rate of change of the burn area, and $h$ is the fuel height.

The rate of water release from combustion can be estimated from eq.~\ref{eq:fuel_consumption_rate} as
\begin{equation}
    S_{w,\text{comb}} = \frac{N_F + N_O}{N_F}\frac{N_{\text{H}_2\text{O}}W_{\text{H}_2\text{O}}}{N_{\text{H}_2\text{O}}W_{\text{H}_2\text{O}} + N_{\text{C}\text{O}_2}W_{\text{C}\text{O}_2}} \rho_f \dot{A} h.
\end{equation}
From this, the ratio of water released from fuel moisture evaporation to that from combustion is:
\begin{equation}
    \frac{S_{w,\text{evap}}}{S_{w,\text{comb}}} =\frac{N_F}{N_F + N_O}\frac{N_{\text{H}_2\text{O}}W_{\text{H}_2\text{O}} + N_{\text{C}\text{O}_2}W_{\text{C}\text{O}_2}}{N_{\text{H}_2\text{O}}W_{\text{H}_2\text{O}}} \mathrm{MC}.
\end{equation}
With a fuel moisture content (MC) of 30\%, this ratio is approximately 54\%, which in order of magnitude is consistent with the simulation result of 33\%, with the difference attributable to the pre-heating and evaporation that occurs ahead of the flame front in the full simulation. For cases at the dry-fuel condition ($\text{MC}=0.01\%$), this ratio is approximately $1.8\times 10^{-4}$, which is negligible, matching with our simulation results.

Similarly, the power released by combustion and consumed by fuel moisture vaporization can be estimated as
\begin{eqnarray}
% \begin{align}
    Q_\text{comb} &=& (H_f - c_{p,f} T_\text{pyr} N_F) \rho_f \dot{A} h,  \\
    Q_\text{evap} &=& (H_w + c_{p,w} T_\text{vap}) \rho_m \dot{A} h.
% \end{align}
\end{eqnarray}
The ratio of fire power consumed by evaporating the fuel moisture can be estimated as
\begin{equation}
    \frac{Q_\text{evap}}{Q_\text{comb}} = \frac{H_w + c_{p,w} T_\text{vap}}{H_f - c_{p,f} T_\text{pyr} N_F} \mathrm{MC}.
\end{equation}
% With 30\% MC, $H_w=2.2564\text{ MJ}$, $H_f = 19.6\text{ MJ}$, $c_{p,w}=4.182\text{ kJ/kg/K}$, $c_{p,f}=1.85\text{ kJ/kg/K}$, $T_\text{vap}=373\text{ K}$, and $T_\text{pyro}=600\text{ K}$, $Q_\text{evap}/Q_\text{comb}=6\%$, which is close to what we observed in the simulation.
For 30\% MC, this gives a ratio of 6\%, closely matching the 2.2\% observed in the full simulation (Figs.~\ref{fig:energy_budget}C and~\ref{fig:energy_budget_ext}), confirming that fuel moisture acts as an energy sink. For cases at the dry-fuel condition ($\text{MC}=0.01\%$), this ratio is approximately $2\times 10^{-5}$, which is negligible and agreeing with our simulation results.

% If your supplement is very short you might need to uncomment the following line to avoid
% layout problems with the figures and tables.
%\newpage

%%%%%%%%%%%%%%%% SUPPLEMENTARY FIGURES %%%%%%%%%%%%%%%

\begin{figure}
  \centering
  \includegraphics[width=0.9\textwidth]{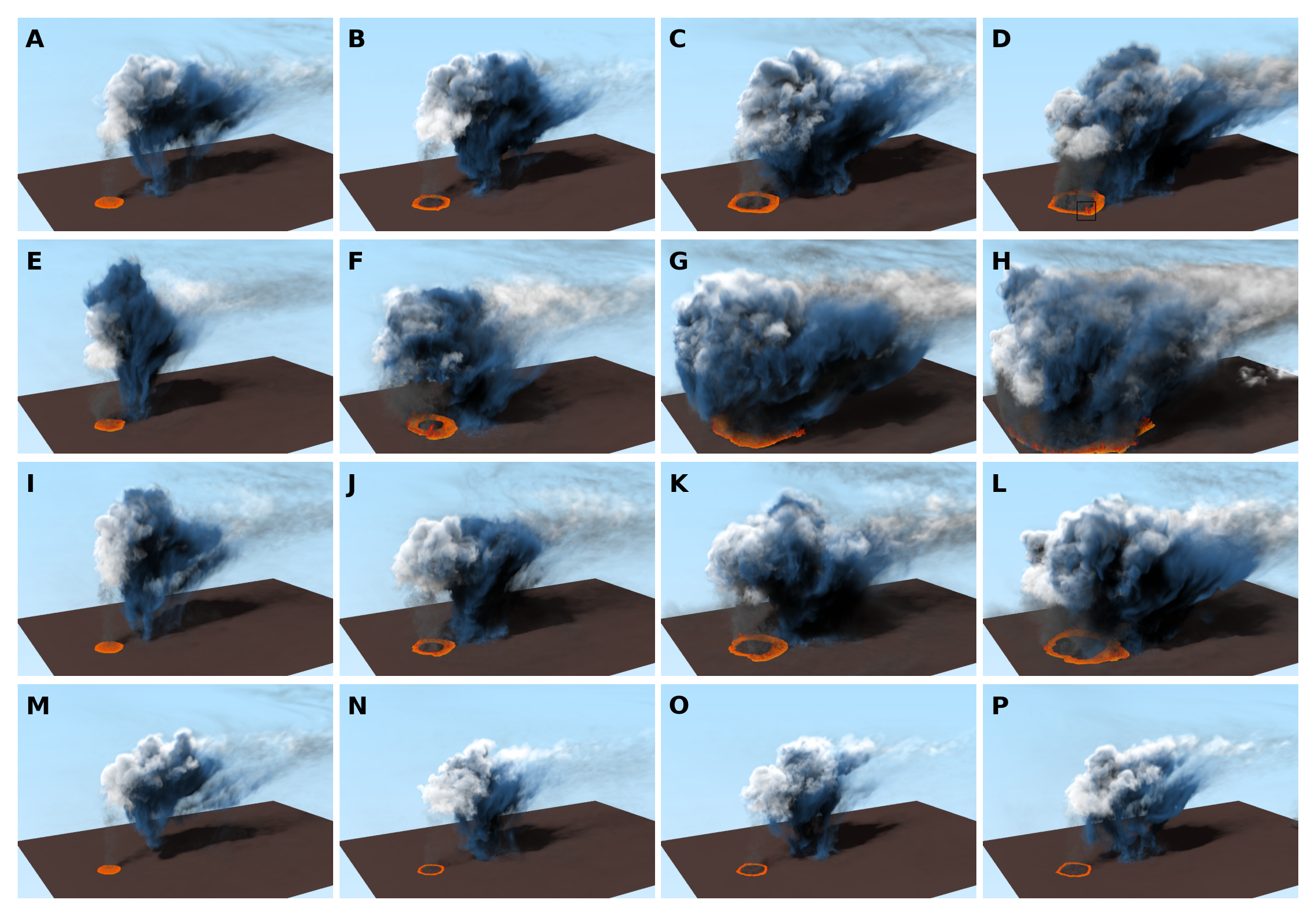}
  \caption{\textbf{3D volume rendering of the fire and pyroCb.} Each row corresponds to a simulation configuration, which are the baseline (\textbf{A} to \textbf{D}), 1/3 wind speed (\textbf{E} to \textbf{H}), 2/3 wind speed (\textbf{I} to \textbf{L}), and 30\% MC (\textbf{M} to \textbf{P}) from top to bottom. Columns from left to right are at time 30, 60, 90, and 120 minutes after ignition.}
  \label{fig:3d_rendering}
\end{figure}

\begin{figure}
  \centering
  \includegraphics[width=0.6\textwidth]{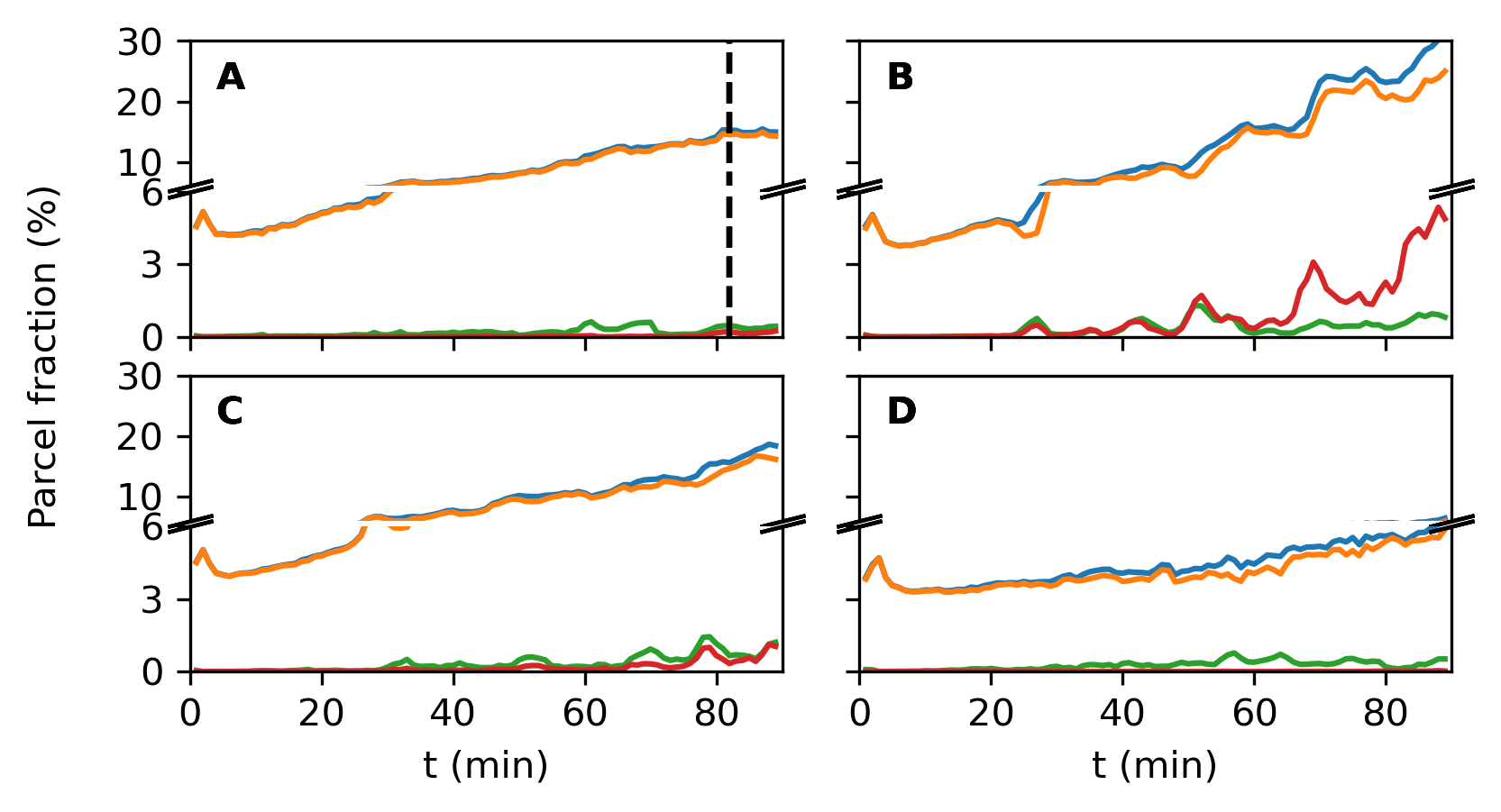}
  \caption{\textbf{The fraction of air parcels that passes through each processes in the conceptural model} (blue: Fire (updraft); orange: Anvil; green: Dispersion; red: Feedback). \textbf{A}, Baseline. The vertical dashed line represent the time stamp at 82 minutes post ignition. \textbf{B}, 1/3 wind. \textbf{C}, 2/3 wind. \textbf{D}, 30\% MC.}
  \label{fig:parcel_count_sankey}
\end{figure}

\begin{figure}
  \centering
  \includegraphics[width=0.9\textwidth]{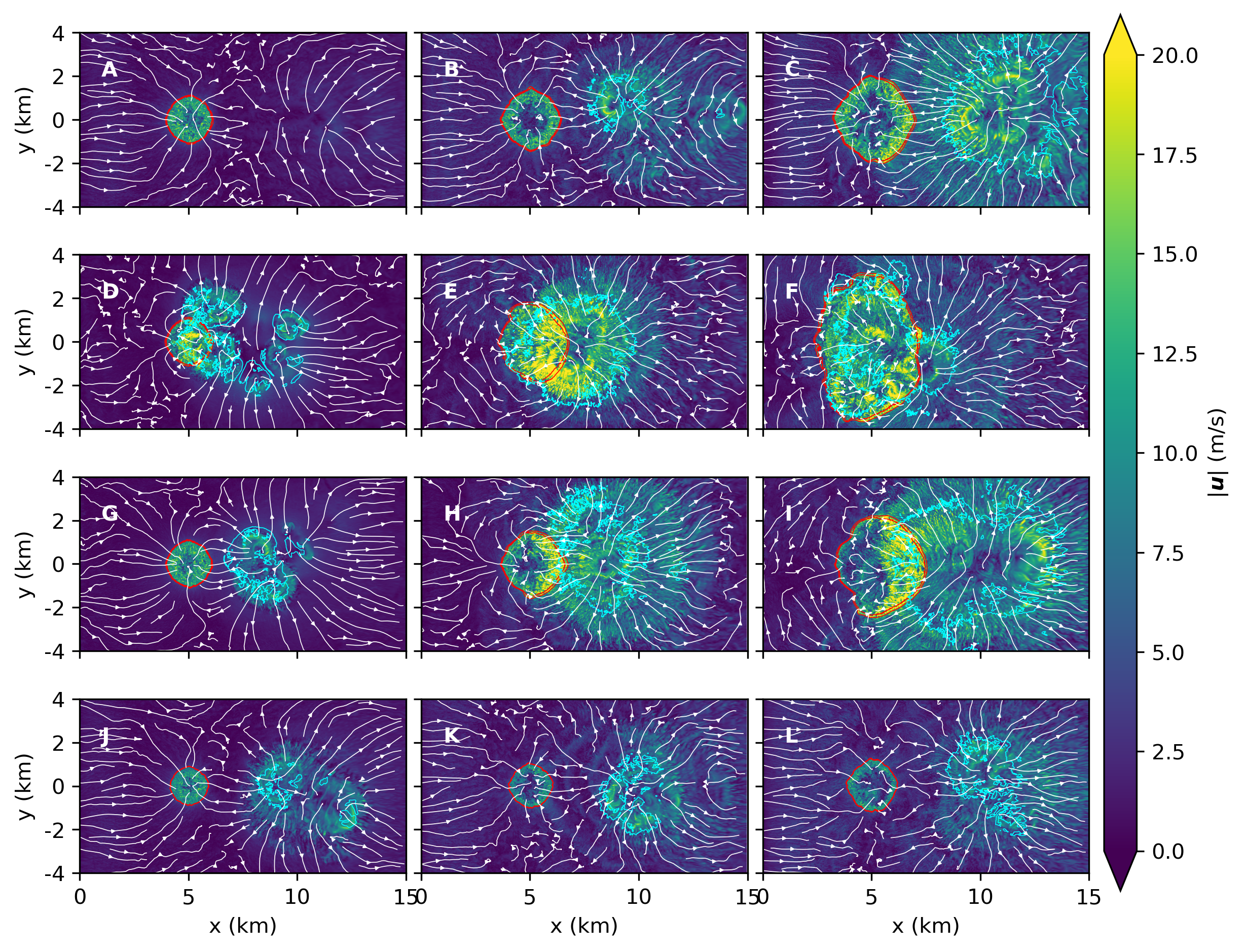}
  \caption{\textbf{Velocity magnitude at 12 m above the ground level with streamlines (white), fire front ($T_s=600$ K, red), and rain front ($q_r=0.01$ g/kg, cyan).} Each row corresponds to a simulation configuration, which are the baseline (\textbf{A} to \textbf{C}), 1/3 wind (\textbf{D} to \textbf{F}), 2/3 wind (\textbf{G} to \textbf{I}), and 30\% MC (\textbf{J} to \textbf{L}) from top to bottom. Columns from left to right are at time 27, 52, and 82 minutes after ignition.}
  \label{fig:ground}
\end{figure}

\begin{figure}[!htb!]
  \centering
  \includegraphics[width=0.9\textwidth]{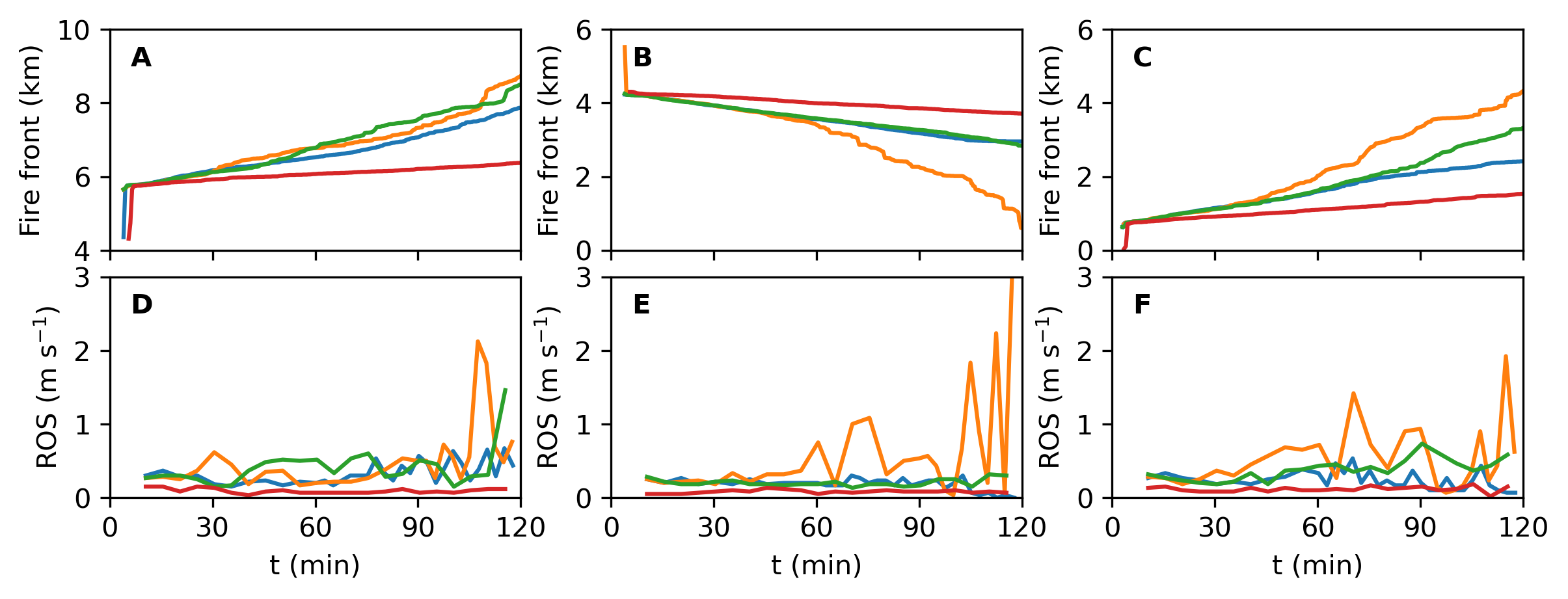}
  \caption{\textbf{The fire front locations and rate of spread (ROS) along different directions}. \textbf{A}, Fire front location of the head fire. \textbf{B}, Fire front location of the back fire. \textbf{C}, Fire front location of the left flank. \textbf{D}, ROS of the head fire. \textbf{E}, ROS of the back fire. \textbf{F}, ROS of the left flank. Legends used by the lines for the four simulation cases are blue: baseline; orange: 1/3 wind; green: 2/3 wind; red: 30\% MC.}
  \label{fig:fire_front_and_ros}
\end{figure}

%%%%%%%%%%%%%%%% SUPPLEMENTARY TABLES %%%%%%%%%%%%%%%

\begin{table}  % Do not use \begin{table*}
  \centering
  % Captions go above tables
  \caption{\textbf{Mean thermodynamic quantities of updraft parcels.} This table presents key thermodynamic variables for mean updraft parcels, calculated from 40 minutes post-ignition and corresponding to the thermodynamic plots in fig.~\ref{fig:skew_t_parcel}. The values for fireCAPE (fire-adjusted Convective Available Potential Energy) and $w_\text{max}$ (theoretical maximum vertical velocity) provide a quantitative basis for comparing the convective potential across the four simulation cases, highlighting the significant reduction in available buoyant energy in the 30\% MC case.}
  \label{tab:skew_t_parcel} % give each table a logical label name
  \begin{tabular}{ l|lcccc  }
    \hline
    \multicolumn{2}{c}{}  & Baseline & 30\% MC & 1/3 Wind & 2/3 Wind \\
    \hline
    \multicolumn{2}{c}{fireCAPE [J/kg]} & 2899.07 & 1921.25 & 2568.69 & 2811.64 \\
    \multicolumn{2}{c}{$w_\text{max}$ [m/s]} & 76.15 & 61.99 & 71.68 & 74.99 \\
    \hline
    & Pressure [hPa] & 573.30 & 580.92 & 580.92 & 565.76 \\
    Centerline & Temperature [$C^\circ$] & 2.18 & 1.21 & 2.07 & 1.18 \\
    LCL & Dewpoint [$C^\circ$] & -0.66 & -1.70 & -0.79  & -1.67 \\
    & Mixing ratio [g/kg] & 6.38 & 5.83 & 6.24 & 6.00 \\
    \hline
    Detrainment & Mixing ratio [g/kg] & 0.08 & 0.28 & 0.09 & 0.09 \\
    Height & Relative humidity & 0.47 & 0.78 & 0.49 & 0.51 \\
    \hline
  \end{tabular}
\end{table}

%%%%%%%%%%% CAPTIONS FOR OTHER SUPPLEMENTARY FILES %%%%%%%%%%

\clearpage % Clear all remaining figures and tables then start a new page

\paragraph{Caption for Movie S1.}
\textbf{Animation showing the evolution of the fire and the pyrocumulonimbus cloud for 90 minutes.}
This animation is an animated version of Fig.~\ref{fig:baseline} A, simulates the formation and evolution of a pyrocumulonimbus cloud using the baseline conditions, starting with a concentrated heat source representing a large fire. A white smoke/water vapor plume rapidly ascends due to convection, expanding and drifting with crosswind. Darker, blackish smoke (soot/carbonaceous aerosols) is entrained into the column, illustrating the mixing of water condensate and fire emissions. The expanding shadow on the ground indicates increasing cloud volume and opacity. The video visualizes the dynamic, turbulent nature of these fire-driven weather systems, which inject significant smoke and pollutants high into the atmosphere, sometimes reaching the stratosphere.
Colors represent fire (orange), smoke (grey), cloud (white), and rain (blue).

\paragraph{Caption for Movie S2.}
\textbf{Animation showing the evolution of a fire whirl from 90.75 to 91.75 minutes post ignition in the 1/3 wind simulation.}
Large, violent flames form a distinct vortex, creating a spinning column of fire. Extreme temperatures are shown by vibrant oranges and yellows. Dark particles (embers/soot) spiral upwards in the powerful updraft, illustrating how fire whirls loft burning debris, spreading fires. The fire whirl shown in this animation is located on the north flank of the fire perimeter. Colors represent gas temperature, particle tail lengths represent distance travelled in 250 milliseconds.

\paragraph{Caption for Movie S3.}
\textbf{Animation showing the evolution of the fire and the pyrocumulonimbus cloud for 90 minutes using the 1/3 wind case.}
This animation simulates the formation and evolution of a pyrocumulonimbus cloud using the 1/3 wind case, starting with a concentrated heat source representing a large fire. A white smoke/water vapor plume rapidly ascends due to convection, expanding and drifting with crosswind. Darker, blackish smoke (soot/carbonaceous aerosols) is entrained into the column, illustrating the mixing of water condensate and fire emissions. The expanding shadow on the ground indicates increasing cloud volume and opacity. This video shows repeated interactions of the downdraft with the fire line, leading to rapid intensification. Colors represent fire (orange), smoke (grey), cloud (white), and rain (blue).

%%%%%%%%%%%%%%%% SUPPLEMENTARY REFERENCES %%%%%%%%%%%%%%%

% Do NOT include a reference list in the supplement.
% All references must be in a single list at the end of the main text.
% The copyeditors will ensure that the correct reference list appears with each version of the paper
% (print, HTML, PDF, mobile app, metadata for bibliographic databases etc.)

%TC:endignore

\end{document}